\documentclass[twocolumn,twocolappendix]{aastex631}

\usepackage{amsmath}
\DeclareMathOperator{\sinc}{sinc}

\shorttitle{COMAP Early Science: II. Pathfinder Instrument}
\shortauthors{Lamb et al.}
\graphicspath{{./}{figures/}}
\NewPageAfterKeywords

\begin{document}

\title{COMAP Early Science: II.\ Pathfinder Instrument}

\correspondingauthor{James W.~Lamb}
\email{lamb@caltech.edu}

\author[0000-0002-5959-1285]{James W.~Lamb}
\affil{Owens Valley Radio Observatory, California Institute of Technology, Big Pine, CA 93513, USA}

\author[0000-0002-8214-8265]{Kieran A.~Cleary}
\affil{California Institute of Technology, 1200 E. California Blvd., Pasadena, CA 91125, USA}

\author{David P.~Woody}
\author{Morgan Catha}
\affil{Owens Valley Radio Observatory, California Institute of Technology, Big Pine, CA 93513, USA}

\author[0000-0003-2618-6504]{Dongwoo T.~Chung}
\affil{Canadian Institute for Theoretical Astrophysics, University of Toronto, 60 St. George Street, Toronto, ON M5S 3H8, Canada}
\affil{Dunlap Institute for Astronomy and Astrophysics, University of Toronto, 50 St. George Street, Toronto, ON M5S 3H4, Canada}

\author{Joshua Ott Gundersen}
\affil{Department of Physics, University of Miami, 1320 Campo Sano Avenue, Coral Gables, FL 33146, USA}

\author[0000-0001-7911-5553]{Stuart E.~Harper}
\affil{Jodrell Bank Centre for Astrophysics, Alan Turing Building, Department of Physics and Astronomy, School of Natural Sciences, The University of Manchester, Oxford Road, Manchester, M13 9PL, U.K.}

\author[0000-0001-6159-9174]{Andrew I.~Harris}
\affil{Department of Astronomy, University of Maryland, College Park, MD 20742}

\author{Richard Hobbs}
\affil{Owens Valley Radio Observatory, California Institute of Technology, Big Pine, CA 93513, USA}

\author[0000-0003-3420-7766]{H\aa vard T.~Ihle}
\affil{Institute of Theoretical Astrophysics, University of Oslo, P.O. Box 1029 Blindern, N-0315 Oslo, Norway}

\author{Jonathon Kocz}
\affil{California Institute of Technology, 1200 E. California Blvd., Pasadena, CA 91125, USA}
\affil{Department of Astronomy, University of California, Berkeley, CA, 94720, USA}

\author[0000-0001-5213-6231]{Timothy J.~Pearson}
\affil{California Institute of Technology, 1200 E. California Blvd., Pasadena, CA 91125, USA}

\author[0000-0001-7612-2379]{Liju Philip}
\affil{Jet Propulsion Laboratory, California Institute of Technology, 4800 Oak Grove Drive, Pasadena, CA 91109, USA}

\author{Travis W.~Powell}
\affil{Owens Valley Radio Observatory, California Institute of Technology, Big Pine, CA 93513, USA}
\affil{Great Basin Unified Air Pollution Control District, Bishop, CA 93514, USA}

%% Tier 2
\author{Lilian Basoalto}
\affil{CePIA, Departamento de Astronomía, Universidad de Concepción, Chile}

\author[0000-0003-2358-9949]{J.~Richard Bond}
\affiliation{Canadian Institute for Theoretical Astrophysics, University of Toronto, 60 St. George Street, Toronto, ON M5S 3H8, Canada}

\author{Jowita Borowska}
\affil{Institute of Theoretical Astrophysics, University of Oslo, P.O. Box 1029 Blindern, N-0315 Oslo, Norway}

\author[0000-0001-8382-5275]{Patrick C.~Breysse}
\affiliation{Center for Cosmology and Particle Physics, Department of Physics, New York University, 726 Broadway, New York, NY, 10003, USA}

\author[0000-0003-2358-9949]{Sarah E.~Church}
\affil{Kavli Institute for Particle Astrophysics and Cosmology \& Physics Department, Stanford University, Stanford, CA 94305, US}

\author{Clive Dickinson}
\affil{Jodrell Bank Centre for Astrophysics, Alan Turing Building, Department of Physics and Astronomy, School of Natural Sciences, The University of Manchester, Oxford Road, Manchester, M13 9PL, U.K.}

\author[0000-0002-5223-8315]{Delaney A.~Dunne}
\affil{California Institute of Technology, 1200 E. California Blvd., Pasadena, CA 91125, USA}

\author[0000-0003-2332-5281]{Hans Kristian Eriksen}
\affil{Institute of Theoretical Astrophysics, University of Oslo, P.O. Box 1029 Blindern, N-0315 Oslo, Norway}

\author[0000-0001-8896-3159]{Marie Kristine Foss}
\affiliation{Institute of Theoretical Astrophysics, University of Oslo, P.O. Box 1029 Blindern, N-0315 Oslo, Norway}

\author{Todd Gaier}
\affiliation{Jet Propulsion Laboratory, California Institute of Technology, 4800 Oak Grove Drive, Pasadena, CA 91109, USA}

\author[0000-0002-4274-9373]{Junhan Kim} 
\affil{California Institute of Technology, 1200 E. California Blvd., Pasadena, CA 91125, USA}

\author{Charles R.~Lawrence}
\affiliation{Jet Propulsion Laboratory, California Institute of Technology, 4800 Oak Grove Drive, Pasadena, CA 91109, USA}

\author{Jonas G.~S.~Lunde}
\affil{Institute of Theoretical Astrophysics, University of Oslo, P.O. Box 1029 Blindern, N-0315 Oslo, Norway}

\author[0000-0002-8800-5740]{Hamsa Padmanabhan}
\affil{Departement de Physique Théorique, Universite de Genève, 24 Quai Ernest-Ansermet, CH-1211 Genève 4, Switzerland}

\author{Maren Rasmussen}
\affil{Institute of Theoretical Astrophysics, University of Oslo, P.O. Box 1029 Blindern, N-0315 Oslo, Norway}

\author[0000-0001-9152-961X]{Anthony C.~S.~Readhead}
\affil{California Institute of Technology, 1200 E. California Blvd., Pasadena, CA 91125, USA}

\author[0000-0001-5704-271X]{Rodrigo Reeves}
\affil{CePIA, Departamento de Astronomía, Universidad de Concepción, Chile}

\author[0000-0002-1667-3897]{Thomas J.~Rennie}
\affil{Jodrell Bank Centre for Astrophysics, Alan Turing Building, Department of Physics and Astronomy, School of Natural Sciences, The University of Manchester, Oxford Road, Manchester, M13 9PL, U.K.}

\author[0000-0001-5301-1377]{Nils-Ole Stutzer}
\affil{Institute of Theoretical Astrophysics, University of Oslo, P.O. Box 1029 Blindern, N-0315 Oslo, Norway}

\author[0000-0003-0545-4872]{Marco P.~Viero}
\affiliation{California Institute of Technology, 1200 E. California Blvd., Pasadena, CA 91125, USA}

\author[0000-0002-5437-6121]{Duncan J.~Watts}
\author[0000-0003-3821-7275]{Ingunn Kathrine Wehus}
\affil{Institute of Theoretical Astrophysics, University of Oslo, P.O. Box 1029 Blindern, N-0315 Oslo, Norway}

\collaboration{36}{(COMAP Collaboration)}



\begin{abstract}
Line intensity mapping (LIM) is a new technique for tracing the global properties of galaxies over cosmic time. Detection of the very faint signals from redshifted carbon monoxide (CO), a tracer of star formation, pushes the limits of what is feasible with a total-power instrument. The CO Mapping Project (COMAP) Pathfinder is a first-generation instrument aiming to prove the concept and develop the technology for future experiments, as well as delivering early science products. With 19 receiver channels in a hexagonal focal plane arrangement on a 10.4\,m antenna, and an instantaneous 26--34\,GHz frequency range with 2\,MHz resolution, it is ideally suited to measuring  CO($J$=1--0) from $z\sim3$. In this paper we discuss strategies for designing and building the Pathfinder and the challenges that were encountered. The design of the instrument prioritized LIM requirements over those of ancillary science. After a couple of years of operation, the instrument is well understood, and the first year of data is already yielding useful science results. Experience with this Pathfinder will guide the design of the next generations of experiments.
\end{abstract}


%\keywords{Single dish antennas(1460) --- Heterodyne receivers(727) --- Radio astronomy(1338) --- Interdisciplinary astronomy(804)}

%% command after the \subsection to give a symbolic KEY to the https://www.overleaf.com/project/61202f6cad9ccc515efbe6ba

\section{Introduction} \label{sec:intro}

The new field of line intensity mapping (LIM; see \citealt{kovetz_astrophysics_2019} for a review) promises to open new windows on the Universe over a wide range of cosmic epochs. By integrating atomic or molecular line emission from all sources in the line of sight, LIM efficiently affords a view over large volumes of the universe. With this technique, we can probe the evolution of galaxies too faint to be detected using blind pencil-beam surveys.  Line emission from the carbon monoxide (CO) molecule has several useful properties for this purpose. With its strong rotational spectrum at millimeter and shorter wavelengths, CO has long been the object of astrophysical interest for its intrinsic properties, and as a tracer of more common but less easily detected molecules, such as molecular hydrogen. Its prevalence makes it a powerful tool for tracking mass associated with star formation \citep{carilli_cool_2013,tacconi_evolution_2020}. At epochs of interest for early star and galaxy formation redshifted CO emission falls in a spectral window in the centimeter wavelength range that is well suited to ground-based observations. Atmospheric opacity is low, receivers have high sensitivity, radio frequency interference (RFI) is minimal, and appropriate spatial resolution is achieved with 10\,m class antennas.

COMAP seeks to exploit these advantages to trace the global properties of galaxies through the epoch of galaxy assembly and back to the Epoch of Reionization using CO LIM \citep{es_I}. As a first step, the Pathfinder instrument targets the 26--34\,GHz band, where it is sensitive to CO($J$=1--0) (rest frame 115\,GHz) from redshift $z=2.4$--$3.4$, with a fainter contribution from CO($J$=2--1) (rest frame 230\,GHz) at $z=6$--$8$. COMAP aims to constrain the distribution of CO-emitting galaxies, expressed as a power spectrum that is derived from the Fourier transform of the three-dimensional distribution of sky brightness. The domain of the transformed data is the wavenumber, generally written $k$, and the range is the brightness power spectral density.

In brief, the COMAP instrument comprises a 19-feed focal plane array receiver mounted at the secondary focus of a 10.4\,m diameter Cassegrain telescope. The 26--34\,GHz signals are downconverted to 2--10\,GHz for transmission to the antenna sidecab. There they are converted to four 0--2\,GHz bands for digitizing, Fourier-transformation, and integration in dedicated digital hardware. The telescope is located at the Owens Valley Radio Observatory, which has excellent weather for observing at centimeter wavelengths year round, and full local technical support.

Table~\ref{tab:instrparams} lists the main instrumental parameters for COMAP. These are based on available equipment and technology selected to meet the science requirements. Fortuitous availability of a Leighton 10.4\,m antenna \citep{leighton_final_1977} from the decommissioned CARMA array \citep{woody_carma:_2004}, prescribed some constraints, but compromises were minimal, especially compared with the advantage of having an existing highly performant, well characterized antenna.
\begin{deluxetable}{lll}[b]
\tabletypesize{\scriptsize}
\tablecaption{Principal COMAP instrument parameters. \label{tab:instrparams}}
\tablehead{
    \colhead{Parameter} & \colhead{Value} & \colhead{Comment} 
} 
\startdata 
Antenna diameter & 10.4\,m & Existing antenna \\
Angular resolution & 4\arcmin.5 FWHM & $k_{\perp} \lesssim$ 0.76\,Mpc$^{-1}$ \\
Scan size & $\sim$\,2\degr & $k_{\perp} \gtrsim$ 0.028\,Mpc$^{-1}$ \\
System temperature & 44\,K (median) & Current technology\\
Feeds & 19 & Hexagonal array; \\
Polarization & Left-circular & Standing wave control\\
Frequency range & 26--34\,GHz & $k_{\parallel} \gtrsim$ 0.006\,Mpc$^{-1}$ \\
Native freq. resol. & {1.95\,MHz} & Extended science goals\\
Science freq. resol. & 31.25\,MHz & $k_{\parallel} \lesssim$ 1.5\,Mpc$^{-1}$ \\
Scan rate\tablenotemark{{\footnotesize a}} & 1\degr\,s$^{-1}$ & Antenna drive limited \\
Integ. time\tablenotemark{{\footnotesize b}} & 20\,ms & Several samples per \\
& & beamwidth\\
\enddata
\tablenotetext{a}{A high scan rate is desirable so that the time to scan the source is small compared to changes in the atmosphere and amplifier gains.}
\tablenotetext{b}{This is the fundamental integration time implemented in the spectrometer}
\tablecomments{Instrument parameters were based on a combination of the science requirements and properties of existing components.}
\end{deluxetable}

Theoretical intensity modeling predictions vary widely and are regularly updated \citep{es_V}, but for the purposes of designing the COMAP Pathfinder, a canonical emission spectrum was assumed to be a distribution of sources of about 40\,\textmu K brightness temperature, with a spectral width for an unresolved galaxy of $\sim$\,40\,MHz \citep{li_connecting_2016}. With a system temperature of $\sim$\,40\,K, the signal-to-noise ratio is of order $10^{-6}$. For a spectral channel of width $\Delta f$, the standard deviation normalized to the mean for an integration time $\tau$ is $\sigma = 1/\sqrt{\Delta f \tau}$. Assuming $\Delta f = 40$\,MHz, achieving $\sigma = 3 \times 10^{-7}$ requires an integration time of $\tau \approx 3$\, days per pixel. A 1\,deg$^{2}$ patch with 5\arcmin~resolution would then need about 1.3\,years observing time. Multi-feed receivers (see, e.g., \cite{emerson_multi-feed_1995}) reduce the integration time in proportion to the number of feeds. A practical configuration of 19 feeds in a hexagonal close-packed array was adopted for COMAP, reducing the time for this idealized observation to about 24 days of integration time. Multiple feeds also provide extra degrees of freedom to mitigate some systematic errors.

\section{Instrument Overview} \label{sec:overview}

The selection of 19 feeds was a compromise between integration speed and cost. Moreover, the speedup is less than proportional to the number of feeds since aberrations and blockage start to become significant limitations.

Coverage in $k_\perp$-space is determined by angular resolution and scan extent, while the $k_\parallel$-space range is defined by the frequency resolution and span. Table~\ref{tab:instrparams} shows the $k$ ranges defined for the COMAP instrument. These fit well with the target $k$-range of at least 0.05--0.5\,Mpc$^{-1}$, a reasonable choice based on current modeling \citep{es_V}. The scan extent is not limited by the antenna, but by covering a small field long enough to achieve a statistically significant measurement based on the fiducial model. In the spectral domain the resolution does not need to be much narrower than the galaxy or galaxy cluster linewidths, but other science, such as radio recombination line detection, is enabled by the 1.95\,MHz resolution.

The following sections provide an overview of the design and implementation of the COMAP instrument. In the design details, we focus on aspects that are particularly challenging for LIM. These are viewed in the context of general considerations of the stability of subsystems over time scales of interest. We present measurements of components, subsystems, and finally the overall performance of the system based on observations of blank sky and astronomical objects such as Jupiter. In the discussion section we use the the experience gained to extrapolate to future instruments.

In designing the COMAP system it was possible to focus on the challenges posed by LIM without regard for requirements for more general observations. Stability of the passband was one of the most critical issues, and any potential perturbing effects, particularly standing waves, were evaluated and mitigated as much as possible. Adverse influences are primarily temperature changes and cable flexing. In evaluating the importance of standing waves, we need to know the ripple amplitude in the spectrum relative to the nominal power, $P_\mathrm{r}$, the mean frequency in the spectrum, $f_\mathrm{mean}$, the integration time per pixel, $t_\mathrm{int}$, the scan time $t_\mathrm{scan}$, and the resolution bandwidth, $\Delta f$. In Appendix \ref{subsec:sw}, we derive a criterion that ensures that the rms of the \textit{change} in ripple over the scan time is less than the statistical noise if
\begin{equation}\label{eq:rateofchange1}
\frac{\mathrm {d} \tau} {\mathrm {d} t} \ll \frac {2}{\pi f_\mathrm{mean} P_{\mathrm {r}} t_\mathrm{scan} t_{\mathrm{int}}^{1/2} \Delta f^{1/2}},
\end{equation}
where $\tau$ is the change in the two-way transit time in the standing wave ``cavity''. For cases where this is thermally driven, this can be used to bound the acceptable temperature change rate, $\mathrm{d}T/\mathrm{d}t$.

Spillover and sidelobes falling on the ground or bright celestial objects are other potentially detrimental effects. Strategies to avoid these phenomena are discussed in the relevant component or system sections. Details of how these effects were evaluated are given in the Appendix.

We will outline the general physical layout of COMAP and the signal flow from the antenna to the spectrometers, before presenting with more detailed descriptions of constituent components.

\subsection{Telescope Layout} \label{subsec:tellayout}

\begin{figure}[b!]
\epsscale{1.15}
\plotone{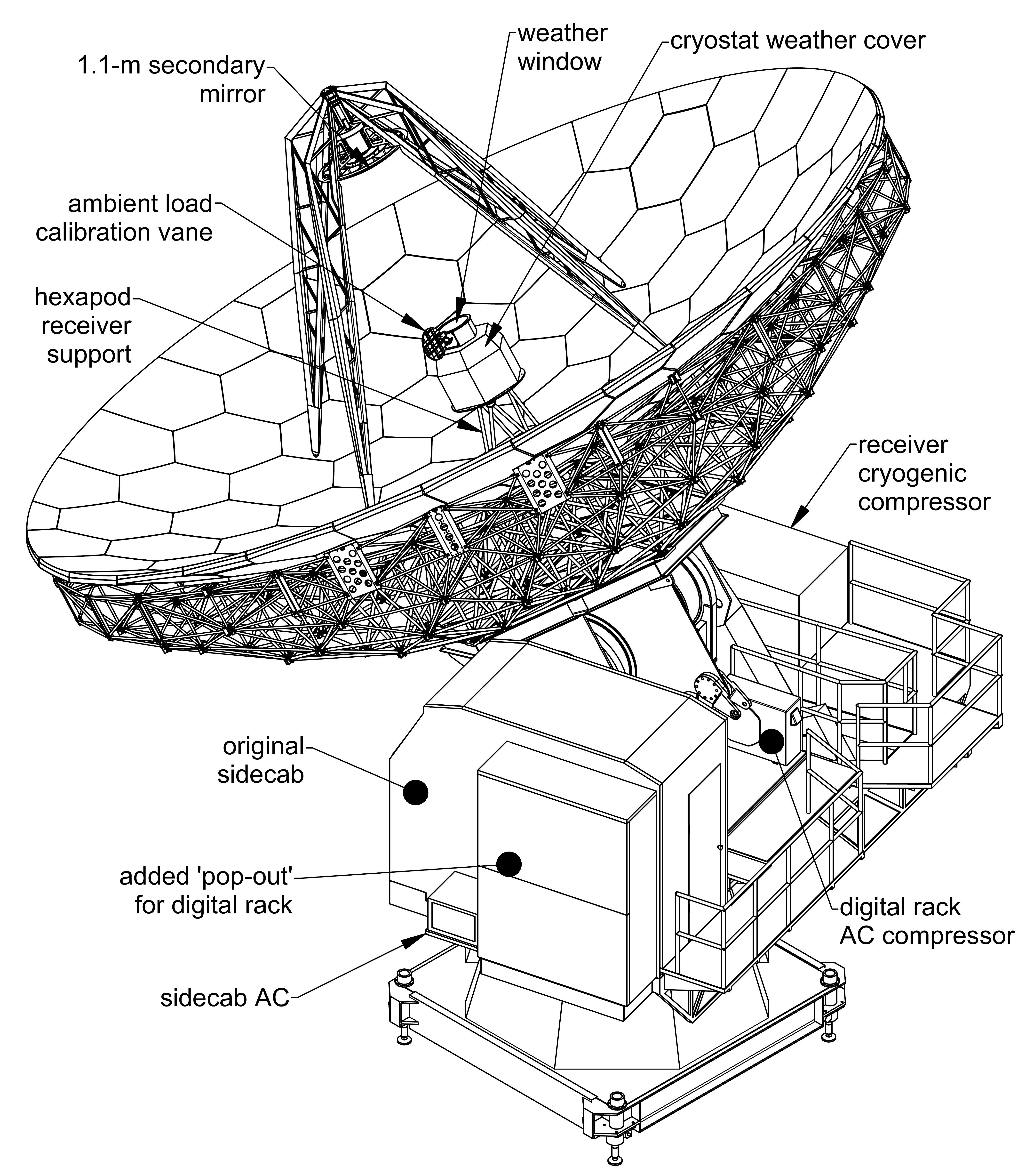}
\caption{Main components of the COMAP telescope. A new secondary mirror was fabricated so that the focal plane receiver could be placed in front of the primary (see text for details). Another major modification was the ``pop-out'' extension to the sidecab for the digital electronics. The two air conditioning units critical to performance are indicated.} \label{fig:antenna}
\end{figure}

\begin{figure*}[t]
\epsscale{1.15}
\plotone{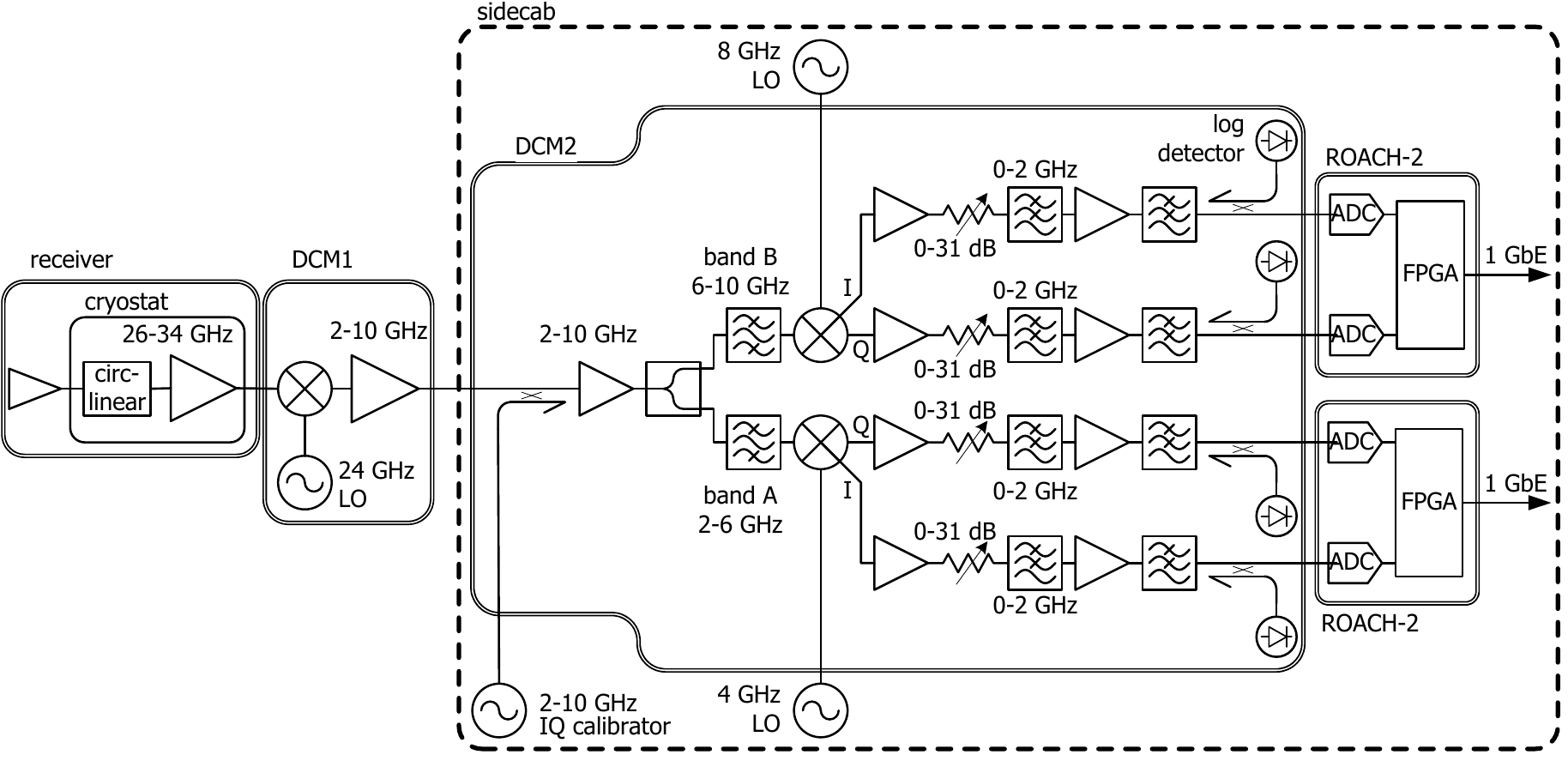}
\caption{Signal path for the COMAP receivers. This path is essentially identical for each of the 19 pixels, except that there is one 24\,GHz LO per DCM1, and one 4\,GHz LO, one 8\,GHz LO and one IQ calibration source common to DCM2s for all pixels.}  \label{fig:signal}
\end{figure*}

Figure \ref{fig:antenna} shows the Leighton 10.4\,m antenna with the receiver package mounted at the secondary focus. Matching the primary to the focal plane array receiver required a new secondary mirror that was larger than the original, as well as having a shorter focal length. An ambient temperature absorber vane is mounted on the receiver package so that the system noise temperature can be periodically calibrated. The receiver package has a cryostat with a mechanical cryocooler for the low-noise amplifiers (LNAs), and its compressor is located on the azimuth platform, connected with flexible helium lines. Signals from the receiver are sent on coaxial cable to the sidecab where the backend electronics are located.

To accommodate the required volume of electronics in the sidecab a ``pop-out'' was added. Most of the pop-out is occupied by an RFI-tight cabinet that contains the digital electronics with its own air conditioning. The top of the pop-out and the remainder of the sidecab is used for analog electronics, antenna control computer and the elevation cable wrap. It uses the original sidecab air conditioning and temperature control. Connections to the central electronics building for frequency references, control and monitoring, and data transfer are on optical fiber.

\subsection{Signal Path} \label{subsec:sigpath}

Figure\,\ref{fig:signal} is a schematic view of the RF path, starting from the feed horns. These are attached to the outside of the cryostat with waveguide vacuum window assemblies. Cooled polarizers convert left-circular polarization to the linear polarization accepted by the waveguide 26--34\,GHz low-noise amplifiers (LNA). Signals from the LNAs are mixed in the first downconverter modules (DCM1) with a 24\,GHz local oscillator (LO), converting them to an intermediate frequency (IF) of 2--10\,GHz. There are four DCM1s, each processing five receiver outputs. As well as the frequency conversion, they amplify the IFs before transmitting them on coaxial cables to the sidecab.

In the sidecab, the signal from each feed is again amplified, and split into \textit{band A} (2--6\,GHz) and \textit{band B} (6--10\,GHz). Each band goes to a second downconverter module (DCM2) with an IQ mixer fed by a local oscillator for the relevant band (4\,GHz and 8\,GHz respectively). In a conventional sideband-separating system \citep{maas_microwave_1992}, the in-phase (I) and quadrature (Q) outputs at 0--2\,GHz would be connected to an analog quadrature hybrid that recombines the signals into the lower sideband (LSB), and upper sideband (USB) of the relevant LO. Cleanly separating the sidebands demands extremely good phase and amplitude balance, which is difficult to achieve over a wide bandwidth. A digital quadrature combination is significantly more accurate \citep{finger_calibrated_2013}. The I and Q signals are directly digitized in ROACH-2 spectrometers (CASPER Collaboration, \citet{parsons_scalable_2008}, Digicom Electronics, Inc.) and converted to complex spectra with a polyphase filterbank \citep{price_spectrometers_2021}. I and Q spectra are combined with complex weights on a channel-by-channel basis to give well separated USB and LSB spectra, and the squared magnitudes are integrated for 20\,ms. Each receiver channel then has four output data streams corresponding to sky frequencies denoted by A:LSB (26--28\,GHz), A:USB (28--30\,GHz), B:LSB (30--32\,GHz), and B:USB (32--34\,GHz), These data streams are captured on a high-speed Ethernet switch and forwarded to the data storage computer.

\vspace{5mm}
\section{Instrument Design Details} \label{sec:details}

\subsection{Antenna} \label{subsec:ant}

One of the Leighton 10.4-m antennas \citep{leighton_final_1977} became available at an opportune time for COMAP. These antennas have been used successfully up to 270\,GHz, guaranteeing excellent performance at 30\,GHz. Significant modifications were required for COMAP to accommodate a focal plane centimeter-wave receiver, including an optical redesign requiring a different secondary mirror and a new location and mount for the receiver. The digital electronics were located in the antenna sidecab, necessitating a modest expansion of its volume. In previous use at the CARMA observatory, the antenna control system was part of a highly integrated software system that was not readily reusable for COMAP. However, it was possible to isolate the antenna drive code running in the antenna and bind it to another control system developed at OVRO.

\subsection{Optical Design} \label{subsec:optics}

\begin{figure}[t!]
\epsscale{1.15}
\plotone{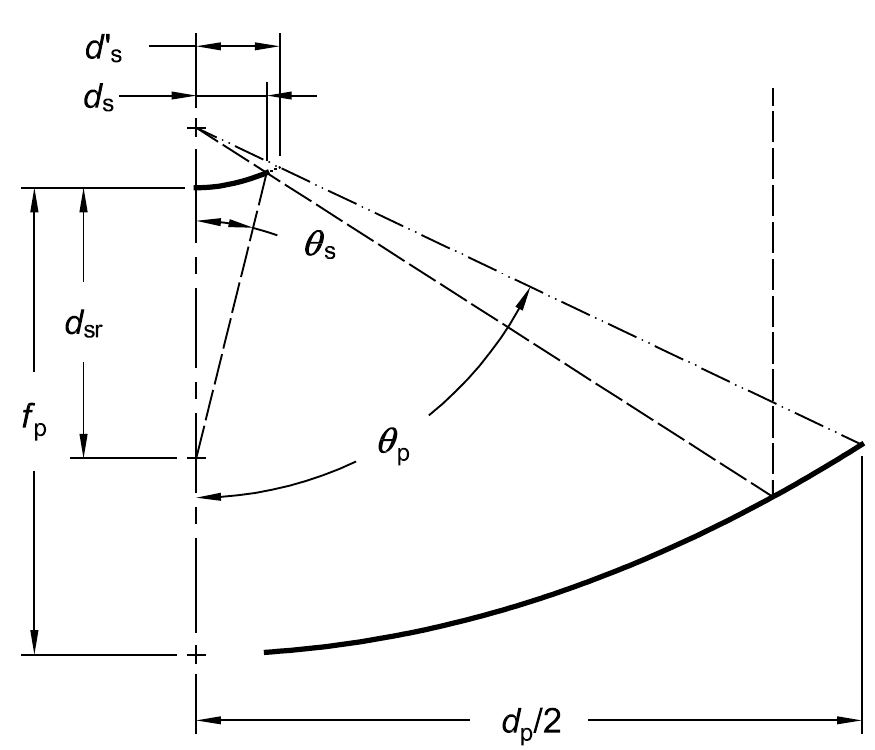}
\caption{Optical design parameters for the COMAP antenna. Values of the parameters are listed in Table \ref{tab:optparams}.\label{fig:antparams}}
\end{figure}

Figure\,\ref{fig:antparams} shows the optical parameters for the COMAP telescope, and the values are listed in Table~\ref{tab:optparams}. Previous receivers were located behind the primary mirror or in the sidecab coupled with a beam waveguide. For a wavelength of 1\,cm the clearance in the backing structure is insufficient for multiple beams to pass through, so the receiver package had to be mounted in front of the surface of the primary, requiring a new secondary mirror with a shorter focal length. This also reduced the required aperture size for the feed horns. To reduce the horn size further the secondary diameter was increased from 600\,mm to 1100\,mm. That diameter is within the size of the aperture at the center of the primary and so incurs no additional blockage penalty.

Choosing a short secondary focal length keeps the feeds and receiver package small, but if it is too short, aberrations are increased, and the blockage of the spherical field between the primary and secondary becomes significant. Additionally, the receiver support becomes unmanageably tall. With the parameters in Table. \ref{tab:optparams}, the receiver is as close to the secondary as possible with out affecting the blockage. 
\begin{deluxetable}{lcDc}
\tabletypesize{\scriptsize}
\tablewidth{0pt} 
\tablecaption{Antenna optical parameters \label{tab:optparams}}
\tablehead{
\colhead{Parameter} & \colhead{Symbol} & \multicolumn2c{Value} & \colhead{Units} 
} 
\decimals
\startdata 
Primary mirror diameter & $d_\mathrm{p}$ & 10.400 & m \\
Primary focal length & $f_\mathrm{p}$ & 4.123258 & m \\
Primary focal ratio & $f_\mathrm{p}/d_\mathrm{p}$ & 0.396 & \\
Secondary to receiver distance & $d_\mathrm{sr}$ & 2.114 & m \\
Half angle subtended by primary & $\theta_\mathrm{p}$ & 64.469 & \degr \\
Magnification & $M$ & 4.50 &  \\
Effective focal length & $f_\mathrm{s} = Mf$ & 18.555 & m \\
Secondary eccentricity & $e$ & 1.571429 & \\
Full secondary mirror diameter\tablenotemark{{\footnotesize a}} & $d'_\mathrm{s}$ & 1.300 & m \\
Actual secondary mirror diameter\tablenotemark{{\footnotesize a}} & $d_\mathrm{s}$ & 1.100 & m \\
Half angle subtended by secondary & $\theta_\mathrm{s}$ & 13.847 & \degr \\
Secondary interfocal distance & $f_\mathrm{c}$ & 2.58426 & m \\
Hyperbola parameters & $a$ & 0.82227 & m \\
{} & $b$ & 1.20823 & m \\
{} & $c$ & 1.29213 & m \\
\enddata
\tablenotetext{a}{Full secondary mirror defines geometry. Actual secondary is undersize to reduce spillover round primary.}
\tablecomments{Values for the optical design. See also Fig.~\ref{fig:antparams}}
\end{deluxetable}

Increasing the diameter of the secondary also reduces spillover from edge diffraction, which varies as the inverse square root of the diameter in wavelengths. By undersizing the secondary diameter by 200\,mm some of the diffracted wave falls on the primary and is reflected to the sky, further reducing ground spillover. Some slight difference in the spillover among feeds will result from differences in the feed offset from the optical axis.

Manufacturing of the secondary mirror was done in an external machine shop on a numerically controlled mill from a billet of 6061 alloy aluminum. The rear surface had pockets milled to reduce the weight without compromising rigidity, and the final mass was 29.2\,kg. The hyperbolic surface has a thickness of 3\,mm.
\begin{figure*}
     \centering
         \includegraphics[width=0.65\textwidth]{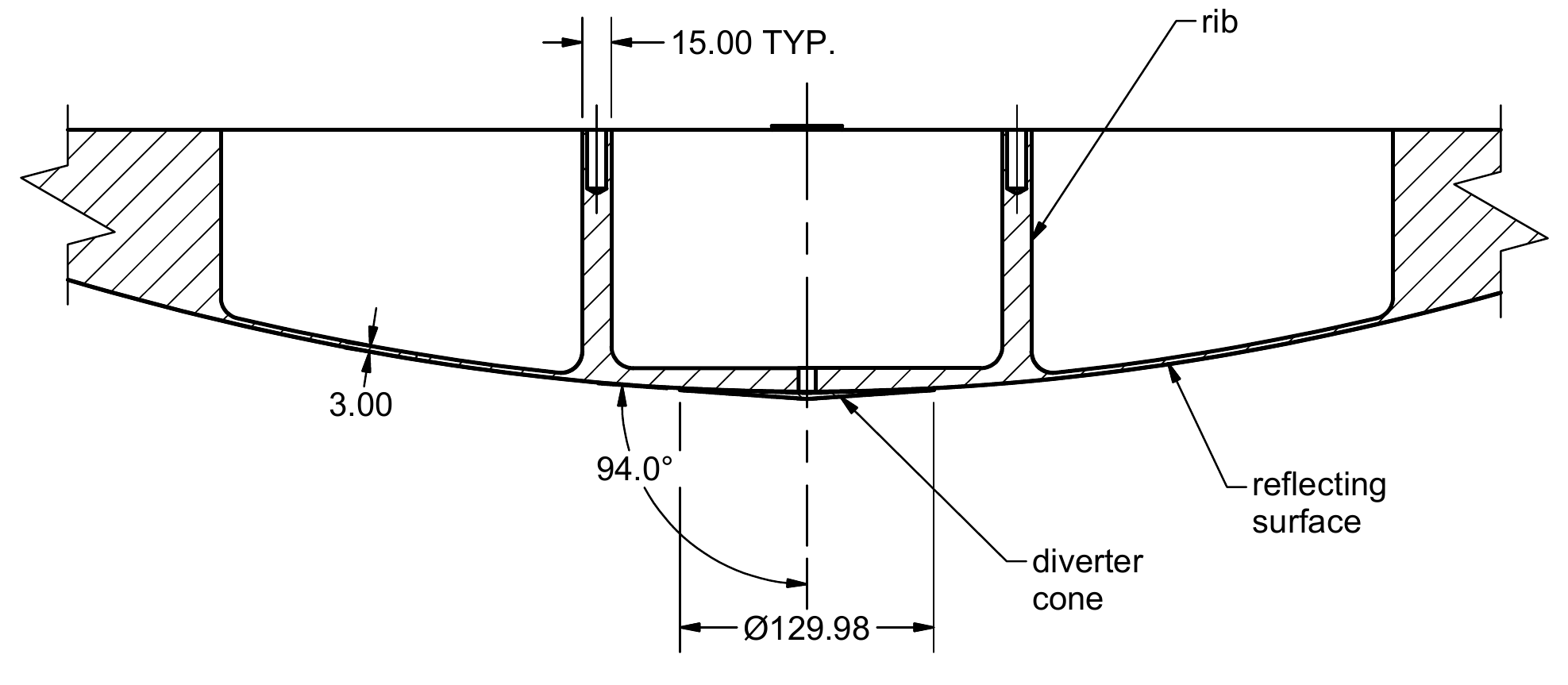}
         \vspace*{-20mm}
         \label{fig:seccone(a)}
         \includegraphics[width=\textwidth]{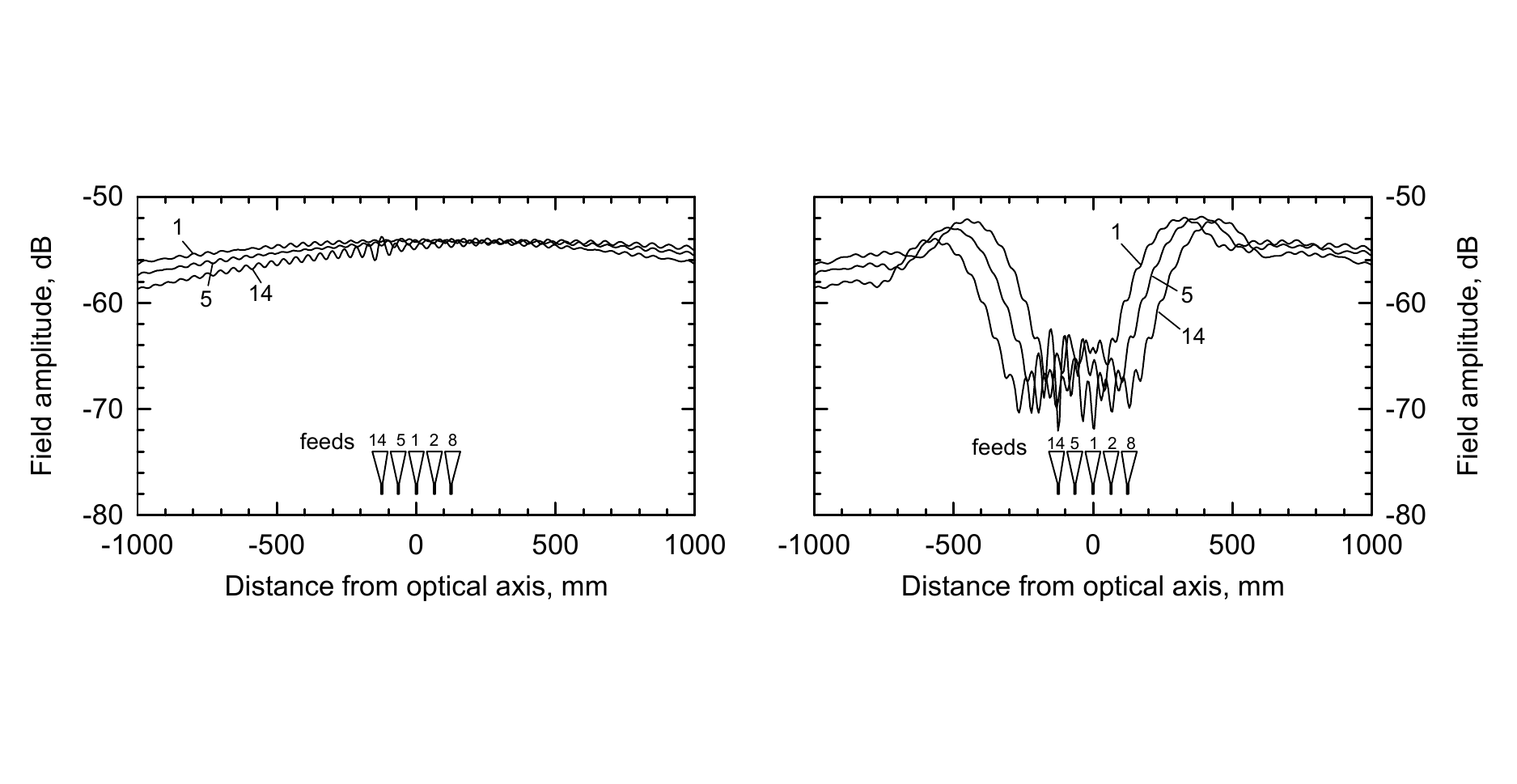}
         \vspace*{-25mm}
        \label{fig:seccone(b)}
        \caption{(\textit{Top}) Cross-section of center of secondary mirror, showing the diverter cone designed to reduce the optical VSWR. Dimensions are in millimeters. (\textit{Bottom}) Calculated field amplitude at 30\,GHz in the feed aperture plane after reflection by the secondary mirror with the diverter cone, (\textit{Left}) no diverter cone and, (\textit{Right}) with diverter cone. The fields from representative feeds (1, 5, and 14, see Figure\,\ref{fig:feedlayout}) are plotted. The feeds drawn to scale at the bottom of the plot illustrate what fraction of the reflected field is captured from the same or a different feed, indicating self- or cross-coupling.}
        \label{fig:seccone}
\end{figure*}

As we discuss in \S\,\ref{subsec:sw}, standing waves that cause ripple in the passband are a significant problem, so a diverter cone was added to the center of the secondary mirror to reduce the optical standing wave. The voltage standing wave ratio (VSWR) was optimized with a wave analysis using a closed-form solution \citep{padman_vswr_1991}, and the resulting design verified using TICRA Tools Physical Optics (PO) analysis \citep{ticra_ticra_2021}. The cone, featured in Figure\,\ref{fig:seccone}, is very shallow and approximately tangential to the hyperboloid where they meet. This cone was machined separately from the mirror and has a threaded stem that screws into the center of the secondary.

By calculating the field from the horn when it is scattered back to the horn aperture plane, the effectiveness can be evaluated. Figure\,~\ref{fig:seccone}  shows that the diverter cone reduces reflections by about 10\,dB.

Corrugated horns \citep{clarricoats_corrugated_1984} were selected as feeds since they have excellent match and beam patterns up to at least 40\% bandwidth. The frequency dependence of the beam pattern is controlled by the phase variation across the aperture, which depends on the length and flare angle of the horn \citep{thomas_design_1978}. A small phase deviation gives a diffraction-limited beam with a width proportional to wavelength, and a large phase deviation produces a constant-width beam. While a constant-width beam maintains aperture efficiency over the band, a diffraction-limited feed gives a frequency-independent antenna beam. The latter was deemed more appropriate for COMAP since it gives a similar resolution in $k_{\parallel}$ at all frequencies. Achieving the desired diffraction-limited feed pattern would require an excessively long horn, but a reasonable compromise was found that gave a modest variation of beamwidth with a 220\,mm long feed.

\begin{figure}[ht!]
    \plotone{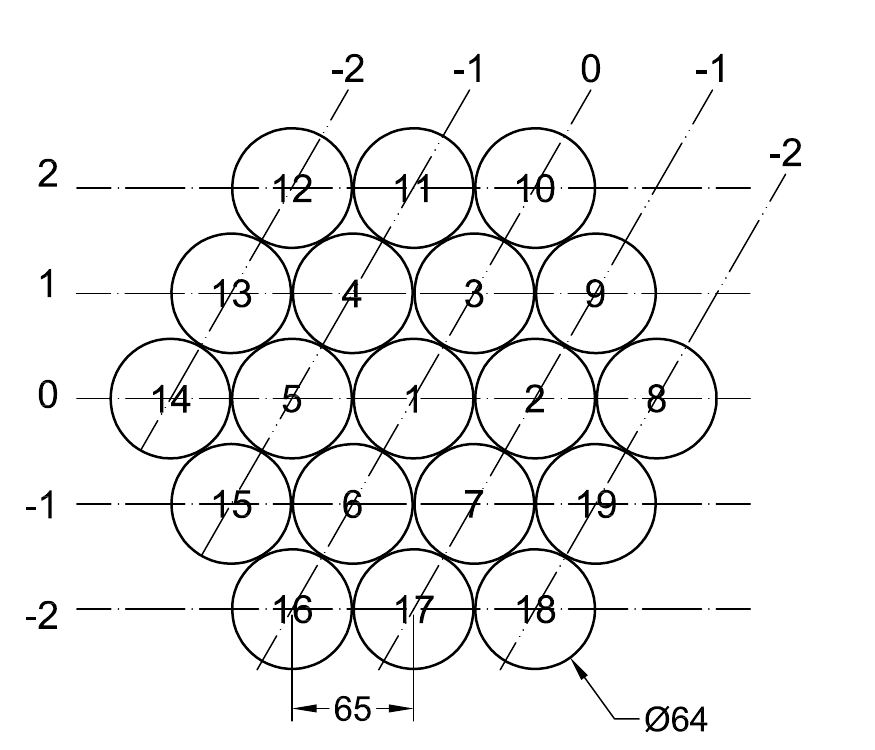}
    \caption{Layout of the feed horns as viewed from the front of the receiver. The numbers at the centers of the feeds are traced through to the final maps, while the indices at the top and side are used for the antenna analysis software. Dimensions for the feed separation, and feed horn outer diameter are indicated in millimeters. In the current installation on the telescope, the receiver was rotated by 90\degr to this orientation.}
    \label{fig:feedlayout}
\end{figure}

Ideally, the feed apertures would be located on the spherical Petzval surface \citep{born_principles_1980}, and pointed near the center of the secondary. However, to simplify the mechanical layout, all the feed apertures were located on a common plane. This has a minimal impact on performance since it has only a small effect on the main beam. The 19 feed apertures are arranged in a hexagonal close-packed array with a spacing of 65\,mm between axes, and 1\,mm between the edges (Figure\, \ref{fig:feedlayout}).

Detailed performance calculations were carried out with TICRA Tools to quantify spillover, aperture efficiency, beam spacing, and sidelobes. Physical Optics (PO) plus Physical Theory of Diffraction (PTD) calculations \citep{clarricoats_handbook_1986} were applied to a detailed model of the telescope, including the panels, secondary with diverter cone, and secondary supports. Beam patterns calculated for the horns using CWGSCAT \citep{hoppe_cwgscat_1993} were used to define the source fields.

\begin{figure}[hb!]
    \centering
    \includegraphics{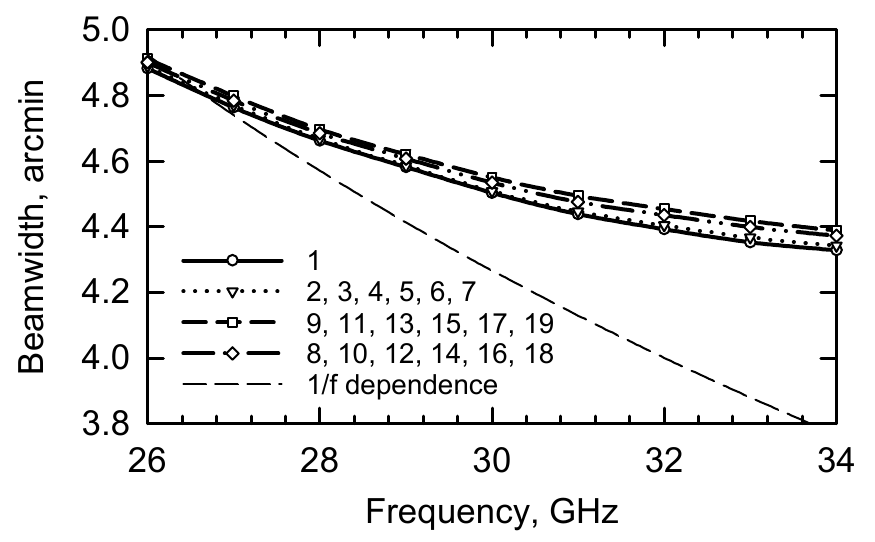}
    \caption{Calculated beamwidth as a function of frequency for all the beams on the sky. The $1/f$ dependence represents the behavior of a diffraction limited beam.}
    \label{fig:beamwidth}
\end{figure}

The horn dimensions selected in combination with the secondary design result in a telescope full-width half-maximum (FWHM) beamwidth of about 4\arcmin.5 and a spacing on the sky of 12\arcmin.0. As shown in Figure \ref{fig:beamwidth}, the beamwidth varies by about 4\% over the observing band, significantly less than would be expected for a diffraction limited aperture antenna.

\begin{figure}[ht!]
    \centering
    \includegraphics{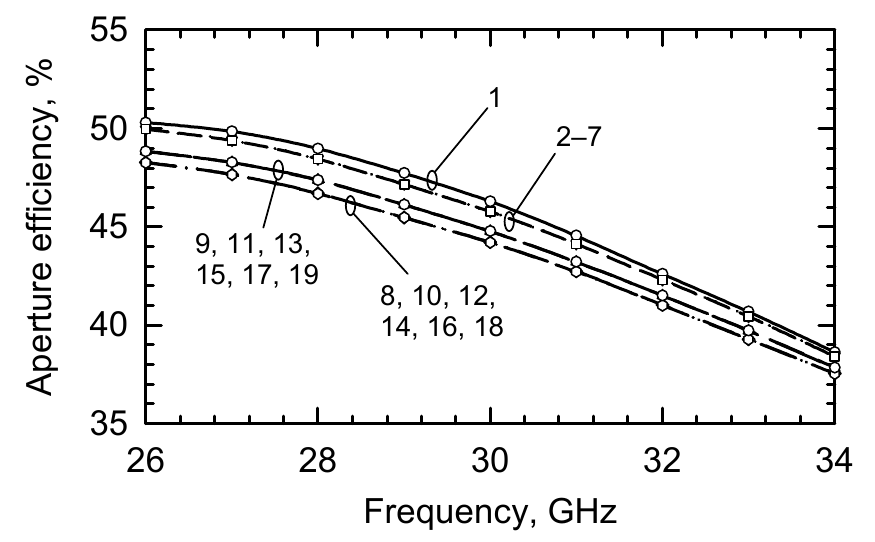}
    \caption{Predicted aperture efficiency, referred to the feed aperture plane. The numbers denote the feeds represented by each set of curves. Although there are only four different offsets from the axis, there are other effects, such as the four-fold feed support symmetry, that make nominally equivalent feeds different. These differences are barely discernible on the graph.}
    \label{fig:apeff}
\end{figure}

Differences in aperture efficiency for the various feed locations are shown in Figure\,\ref{fig:apeff}. These are small, and due mainly to coma. Achieving maximum efficiency requires a frequency-independent illumination of the telescope aperture, so the decrease with frequency is expected and not detrimental for this experiment.

\begin{figure}[hb!]
    \includegraphics{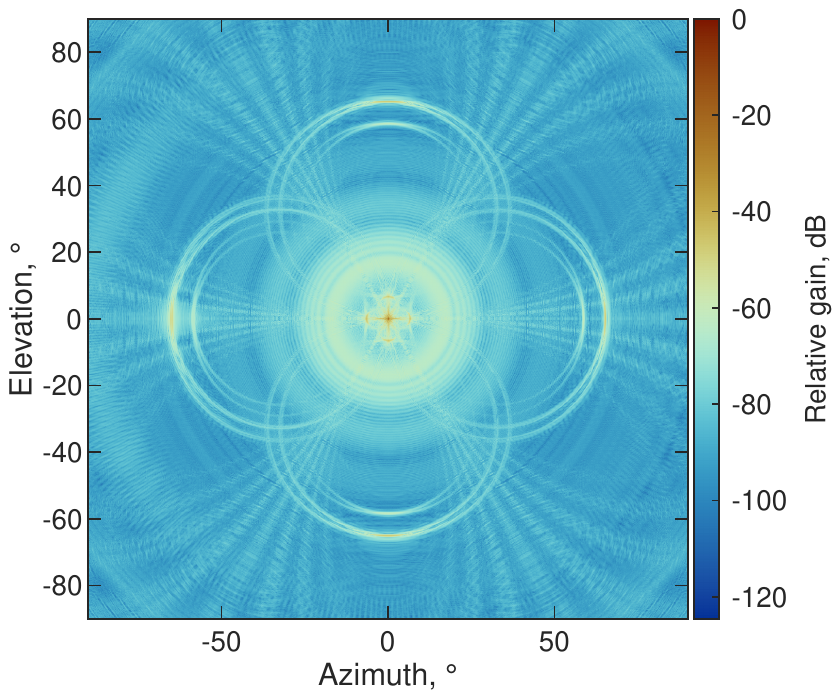}
    \caption{Map of far-out sidelobes. The main beam is about one pixel. Sidelobes are due to diffraction at the aperture, spillover past the secondary, secondary support struts, and panel edges.}
    \label{fig:sidelobes}
\end{figure}

Sidelobes were characterized on a number of scales and at various frequencies using the TICRA Tools package, with a typical map shown in Figure~\ref{fig:sidelobes}. Element sizes for the finite-element analysis were of order a half-wavelength so that the computed field is valid over the forward hemisphere. Limitations are due mainly to some necessary simplifications of the structural elements. For example, the secondary support struts were modeled as two-dimensional elements corresponding to the faces closest to the axis of the dish.

Several sources of sidelobes are identifiable. Diffraction by the antenna aperture produces near-in sidelobes that rapidly fall off from the main beam. Spillover of the feed pattern past the secondary results in an annular sidelobe at about 14\degr \ from boresight. From the work of \citet{rusch_radiation_1982}, the scattering of the outgoing plane wave from the primary by the secondary support struts is predicted to propagate along a conical profile, with the axis of the cone along the strut, and the cone containing the optical axis. This produces annular sidelobes that intersect at the main beam and have a width corresponding to the diffraction size for the length of the strut projected on the aperture plane. Wide-angle rearlobes result from the feed pattern being diffracted at the rim of the secondary and then propagating back past the edge of the primary. Central blockage by the hole in the primary creates a (negative) diffraction pattern similar to the primary, but with a scale size ten times greater and an amplitude of a few percent.

\subsection{Receiver Mount} \label{subsec:rxmount}

To mount the receiver in front of the primary, a hexapod made from aluminum rods was clamped at the lower end to struts in the primary backing structure, extending through the central hole in the primary to the aluminum support plate for the cryostat (Figure\,\ref{fig:antenna}). The triangular configurations match the six-fold symmetries the backup structure is based on, and form a strong support without over-constraint. The orientation was chosen to make the cryocooler cylinder parallel to the elevation axis to minimize potential gravitational effects on temperature stability.

Aluminum struts were selected so that the differential thermal expansion relative to the steel antenna structure would keep the feed-to-secondary distance approximately independent of ambient temperature, minimizing variation of the optical standing wave ripple. For a homogeneous steel structure, the temperature sensitivity of the feed-to-secondary distance would be 25\,$\mu{\text K}^{-1}$. If we apply Equation\,\ref{eq:rateofchange1} with $f_\mathrm{mean}= 30$\,GHz, $P_\mathrm{r}=$\,0.01, $t_\mathrm{scan}=$\,20\,s, $t_\mathrm{int}=$\,100\,ms, and $\Delta f=$\,32\,MHz, and we require the ripple to be an order of magnitude less than the statistical noise, the rate of change of temperature should be less than 36\,mK\,s$^{-1}$. Changes in ambient temperature of this order are expected, though the thermal mass of the struts reduces this somewhat. By combining aluminum struts with the steel a significant improvement is predicted. Since the coefficient of thermal expansion of aluminum is about twice that of steel, and the aluminum components span about 2\,m, while the distance in steel is about 4\,m the net expansion of the feed-to-secondary distance is zero to first order.

A hexagonal shroud fabricated from aluminum sheet, painted white, and lined with 25\,mm thick building foam provides substantial isolation from convective and radiative heat exchange due to ambient temperature and solar radiation. Air is blown into the the shroud through a flexible coaxial aluminum duct from the temperature-controlled sidecab. The diameters of the inner and outer ducts are 150\,mm and 200\,mm, and the space between is filled with insulating foam beads. A second identical duct returns the air to the sidecab. The supply duct also carries the signal cables from the receiver down to the backend electronics.

\subsection{Drives} \label{subsec:drives}

No modifications were made to the existing antenna drive motors or controllers since they had adequate performance. DC motors with brushes are driven by commercial electronic controllers that servo the rate to an applied voltage using tachometers integrated with the motors. Elevation backlash is removed by unbalancing the elevation structure slightly so that the ball-screw drive is always in compression. Azimuth backlash is countered by having two motors with opposite but unequal torque driving a large azimuth gear.

\subsection{Receiver Package} \label{subsec:rxpkg}

\begin{figure*}
    \plotone{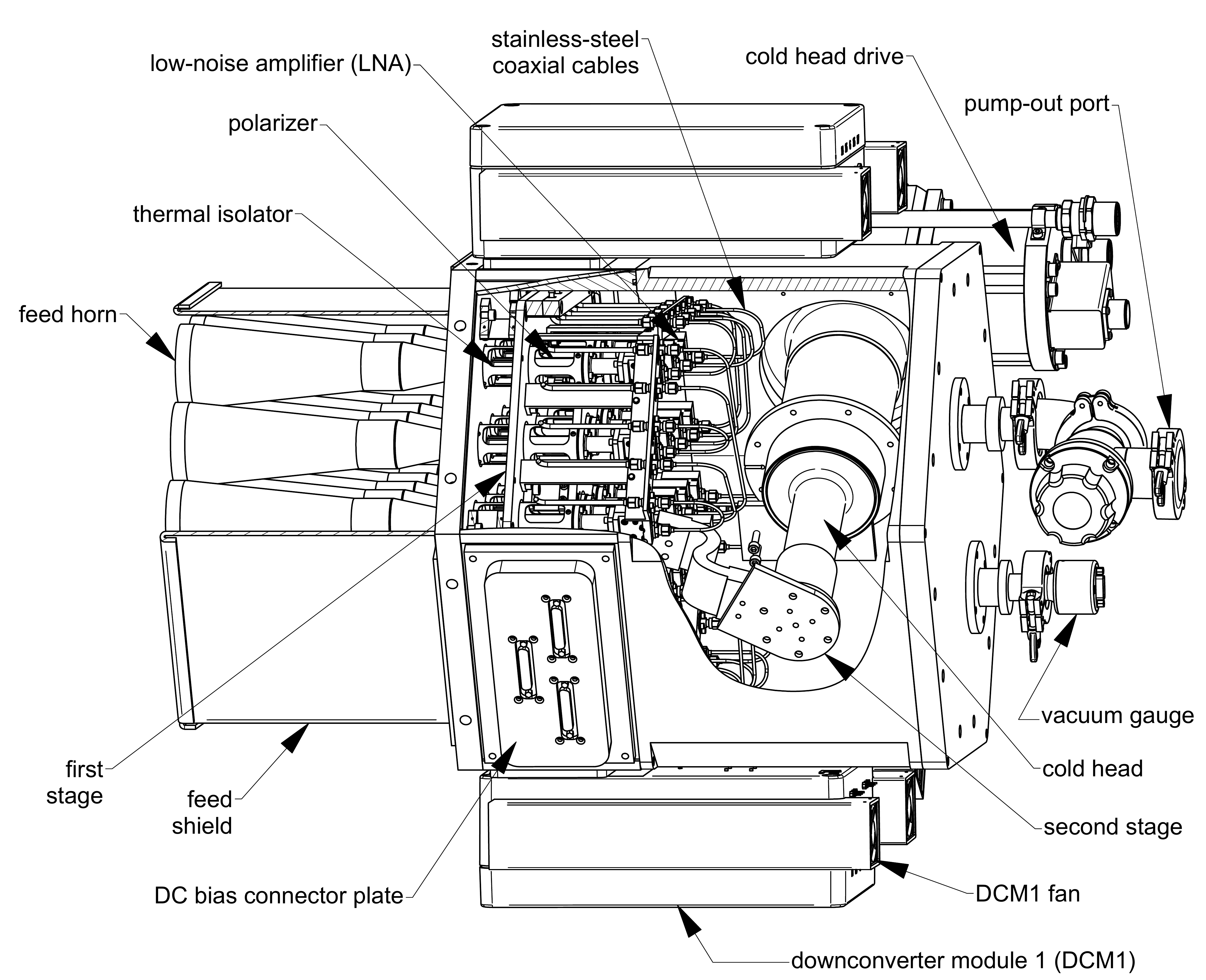}
    \caption{Sectioned view of receiver. The outer vacuum jacket is machined from two aluminum blocks, with aluminum end plates. Details of the waveguide components and the thermal isolation are provided in the text.}
    \label{fig:rx}
\end{figure*}

Nineteen feed horns are mounted on the front of a cryostat (Figure\,\ref{fig:rx}) that contains circular-to-linear polarizers and cryogenic LNAs. DCM1 modules attached to the outside of the cryostat convert the signal from 26--34 GHz down to 2--10 GHz for transmission to the second stage of signal processing in the sidecab. A microwave-transparent window mounted on the top opening of the shroud (\S\,\ref{subsec:rxmount}) protects the receiver.

\subsubsection{Cryostat} \label{subsubsec:cryostat}

Machined from solid aluminum, the main vacuum container of the cryostat is 375 mm high with end plates installed, and 381 mm between flats of its hexagonal cross-section. The top face has mounting holes for the vacuum windows that mate with the feed horns. Inside the vacuum space, an aluminum heat shield is attached to the top plate with G-10 fiberglass epoxy tabs. These are close to the perimeter of the shield with their face normals directed to the cryostat axis so they can flex on cooling while keeping the shield coaxial with the outer container. A gold-plated second stage copper shield is mounted in a similar way inside the first stage shield. The receiver components are individually supported on their respective vacuum window flanges, and thermal connections to the second stage are made with C10100 copper braid, allowing excellent heat transfer with minimal stress (particularly from differential contraction on cooling). Signals from the LNAs are routed using 0.085 stainless steel coaxial cables to a bulkhead on the first stage, and 0.141 stainless steel coaxial cables from there to the vacuum wall vacuum feedthroughs.

Cooling is provided by a Sumitomo RDK-408S2 cold head operating with a CSA-71A compressor. The cold head is inserted through one side of the outer wall of the cryostat, and the first and second stages are attached with multiple sheets of high-purity copper foil to their respective radiation shields.

\subsubsection{Calibration Load} \label{subsubsec:calload}

Following standard millimeter-wave methods \citep{penzias_millimeter-wavelength_1973,ulich_absolute_1976}, amplitude calibration uses an ambient absorber that can be rotated into place in front of the feeds in about a second. The TK RAM absorber (Thomas Keating Ltd) used has a reflection of better than $-25$ dB at 30 GHz. Tessellated 100\,mm square tiles were trimmed to a circular perimeter and mounted on a lightweight frame, and a synchronous AC gear motor inserts and removes it from the beams. With an appropriate geometry for the motor axis, the 432\,mm diameter vane rotates from parallel to the window in the calibration position, to being in the plane of the secondary supports in the ``out'' position (see Figure\,\ref{fig:antenna}). In that orientation it causes no additional shadowing of the aperture.

Thermal insulation foam is glued to the rear to minimize thermal gradients, and both front and back are sprayed with white paint to reduce solar heating. Temperature sensors are embedded in the load and the control system continuously records the value. During early testing it was found that performance was improved by backing the load with aluminum foil. This eliminated the radiation transmitted from the sky and effectively doubled the optical depth of the absorber.

\subsubsection{Waveguide Components} \label{subsubsec:wgcmpts}

\begin{figure}
    \epsscale{1.15}
    \plotone{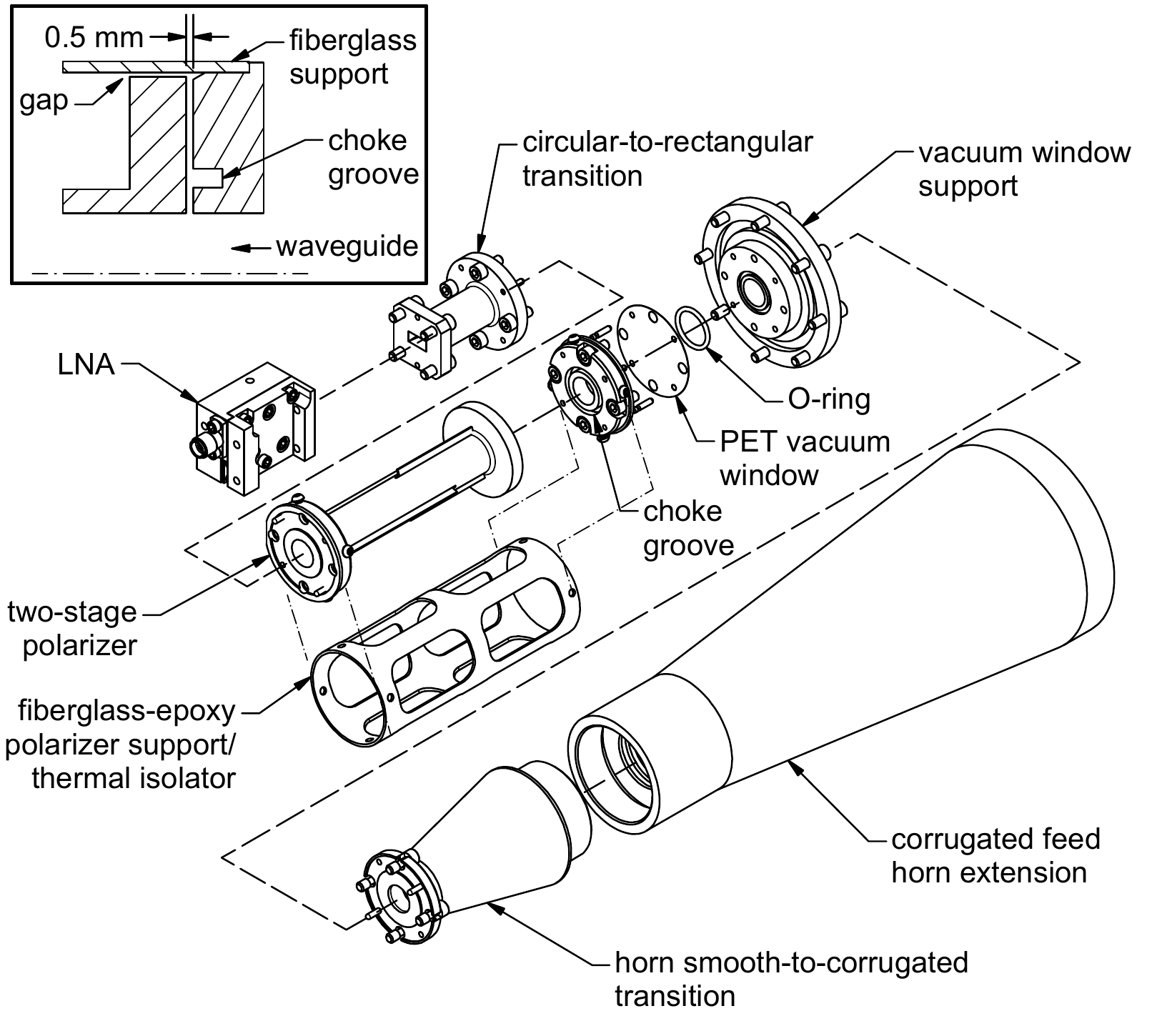}
    \caption{Exploded view of the waveguide components that are located inside the cryostat. The fiberglass sleeve is attached to the polarizer at the left end only, supporting it so that right end is about 0.5\,mm from the choke groove on the window flange (see inset, top left). The window flange is at ambient temperature and captures the vacuum window against the O-ring seal.}
    \label{fig:wgassy}
\end{figure}

\begin{figure}[hb!]
    \epsscale{1.15}
    \plotone{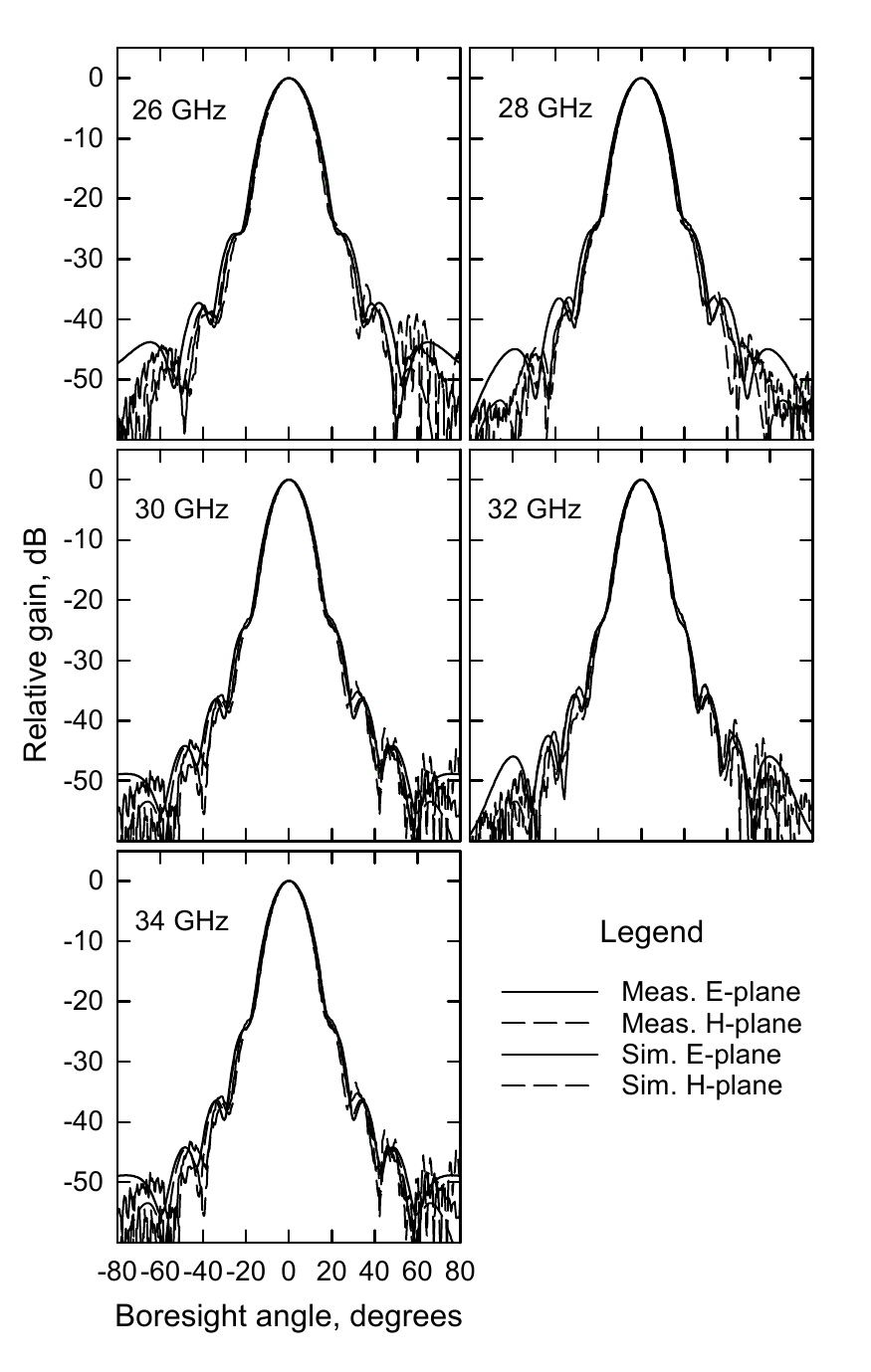}
    \caption{Computed and measured feed patterns. The patterns were measured in an anechoic chamber.}
    \label{fig:feedpatt}
\end{figure}

Figure\,\ref{fig:wgassy} shows the chain of waveguide components in the receiver. The corrugated horn was optimized for beam pattern and return loss using CWGSCAT \citep{hoppe_cwgscat_1993}. Pattern measurements in an anechoic chamber were in excellent accord with the predictions as is evident in Figure\,\ref{fig:feedpatt}.

Since corrugated structures have very low losses \citep{clarricoats_corrugated_1984}, the horn is not cooled. Any slight increase in noise from ambient temperature loss in the horn is offset by avoiding the loss in thicker windows required for cooled horns. This configuration also allows a very small thermal radiation area between ambient and cryogenic temperatures, defined by the area of the waveguide. Vacuum integrity is ensured by a 25\,\textmu m thick DuPont Mylar (PET) window, located on the inside of the vacuum flange allowing feed horns to be removed without breaking vacuum. 

As part of the effort to control ripple in the system, the receivers were designed to use left circular polarization (LCP). Noise emitted by the LNA and reflected by the secondary has its polarization reversed so it does not couple back to the LNA and create ripple in the noise temperature. Similarly, LCP signals from the sky that are reflected by the LNA and secondary are rejected. Right circular polarization from the sky that is reflected at the rectangular waveguide and then by the secondary as LCP couples into the receiver but is uncorrelated with the direct LCP signal and does not produce ripple.

\begin{figure}
    \plotone{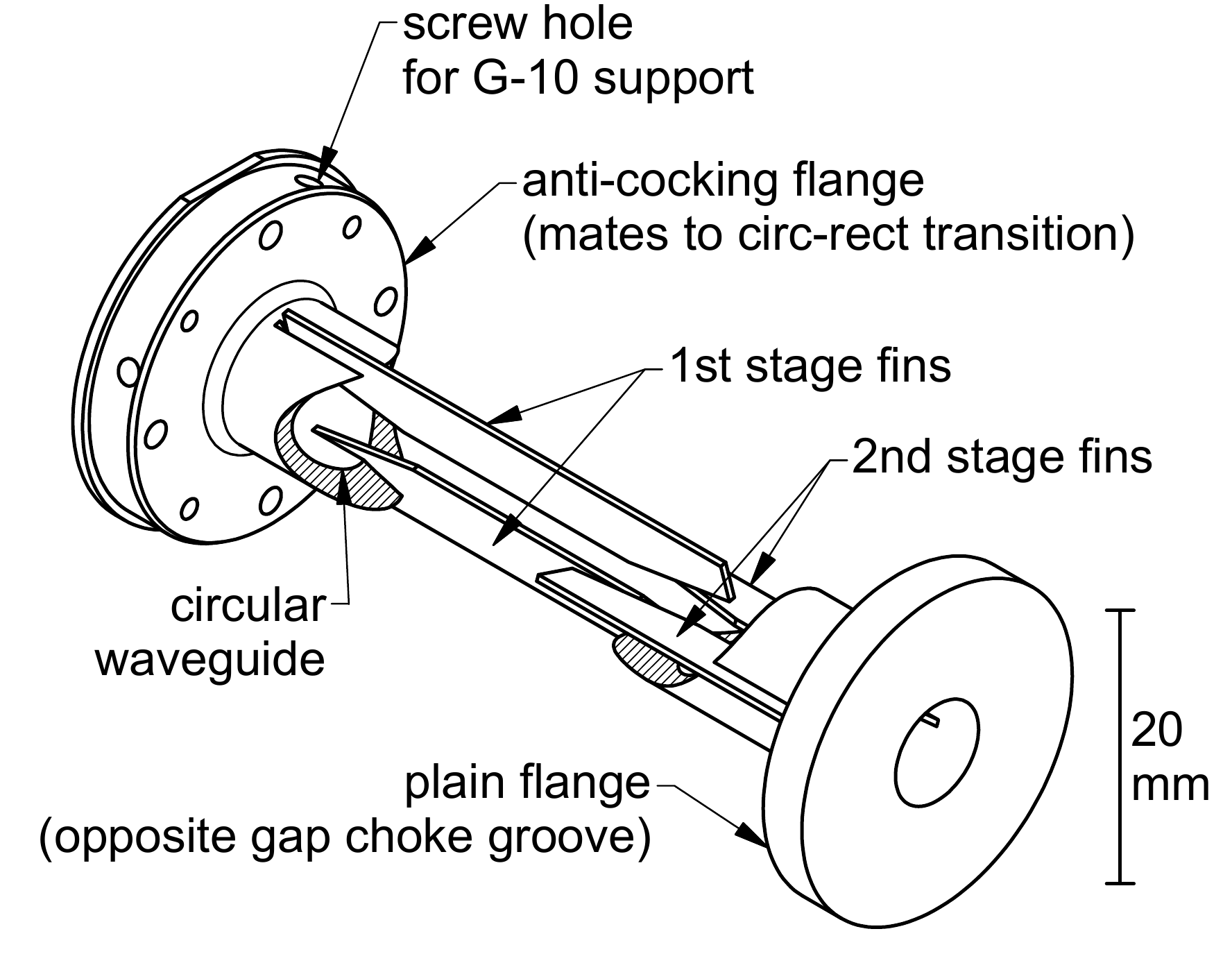}
    \caption{Cut-away view of circular to rectangular polarizer. The vanes that provide the retardation are silver-epoxied into slots in the circular waveguide.}
    \label{fig:polarizer}
\end{figure}

Conversion from circular to linear polarization is effected in a two-stage polarizer, illustrated in Figure\,\ref{fig:polarizer}. The principles of this design are given by \citet{plambeck_circular_2010} based on work by \citet{pancharatnam_achromatic_1955}. It uses waveguide sections that retard one linear polarization relative to the other. The orientation of the principal axes and the retardation per section are chosen to optimize performance across a given band. Increasing the number of stages improves performance but requires increasingly tight tolerances. Electromagnetic modeling indicated that good performance can be achieved with two stages.

Figure\,\ref{fig:polarizer} shows the construction of the polarizers. In each section, retardation of one component of polarization results from two copper fins symmetrically inserted into slots in a brass circular waveguide. The fins are tapered to reduce mismatches, and silver epoxied (EPOTEK H20E) to the waveguide with a jig to ensure alignment. Different lengths and orientations of the fins are used for the two stages. To minimize the overall length, a small separation is used between the stages which requires careful optimization in the simulator to account for inter-stage effects. For corrosion protection and loss reduction a soft-gold plating layer is applied. Verification of the manufactured components was done with two polarizers using a vector network analyzer. This verified that the reflection coefficients were better than $-20$\,dB across the band, and the through signal was found to be essentially independent of the relative rotation of polarizers about their common axis.

As seen in Figure\,\ref{fig:wgassy}, a G-10 fiberglass-epoxy cylinder holds the polarizer at the cryogenic end while the other end is anchored to a flange at ambient temperature. A choke ring \citep{helszajn_passive_1978} on the ambient flange prevents leakage through the gap between it and the polarizer. On cooling, the gap becomes smaller owing to the different thermal coefficients of expansion between the brass polarizer and the fiberglass-epoxy. Simulations confirm that the choke ring operates correctly over the range of gap sizes from ambient to cold. Apertures were machined into the fiberglass-epoxy cylinder to reduce thermal loading. That reduces the stiffness, potentially allowing misalignment between the polarizer and the choked flange so the deformation under gravity was calculated using finite element simulation and the resultant deformed structure was simulated electromagnetically. The effects were determined to be negligible. Measurements of the deflection and transmission of a fabricated device closely matched the simulations.

An electroformed transition was designed to mate the circular polarizer waveguide to the LNA. The mandrel was an aluminum cylinder with flat faces tapered down on four sides to the required rectangular waveguide cross-section at the end. Machining and electroforming were done commercially (A.J. Tuck Co.). Anti-cocking flanges \citep{kerr_waveguide_1999} were used for good integrity of interfaces.

Full electromagnetic simulations of the structure from the LNA input to the horn aperture were carried out with using Ansys HFSS \citep{ansys_ansys_2021} to verify performance. This included the vacuum window, and the waveguide gap choke. Leakage through the gap was verified to be insignificant in terms of cross-talk between feeds.

\subsubsection{Low-noise Amplifiers} \label{subsubsec:lnas}

The front-end amplifiers set the noise performance of the whole receiver provided they have sufficient gain. Amplifiers based on high electron-mobility transistors (HEMT) provide the lowest noise at Pathfinder frequencies. For array applications, LNAs constructed from monolithic microwave integrated circuits (MMIC) can provide consistent performance without the need for extensive fine-tuning. We therefore selected a proven MMIC HEMT-based LNA design for COMAP \citep{tang_full_2006} as the basis for our Pathfinder LNA module. This design, fabricated on Northrop Grumman Corporation's (NGC) 100\,nm InP process, was demonstrated to exhibit noise temperatures in the 7--15\,K range over 26--34\,GHz. A pool of such devices was available at JPL and we also fabricated a new version of this design on NGC's 35\,nm InP node. Both 100\,nm and 35\,nm chips were used interchangeably in the final LNA modules.

\begin{figure}[b]
    \centering
    \includegraphics[width=1.0\linewidth]{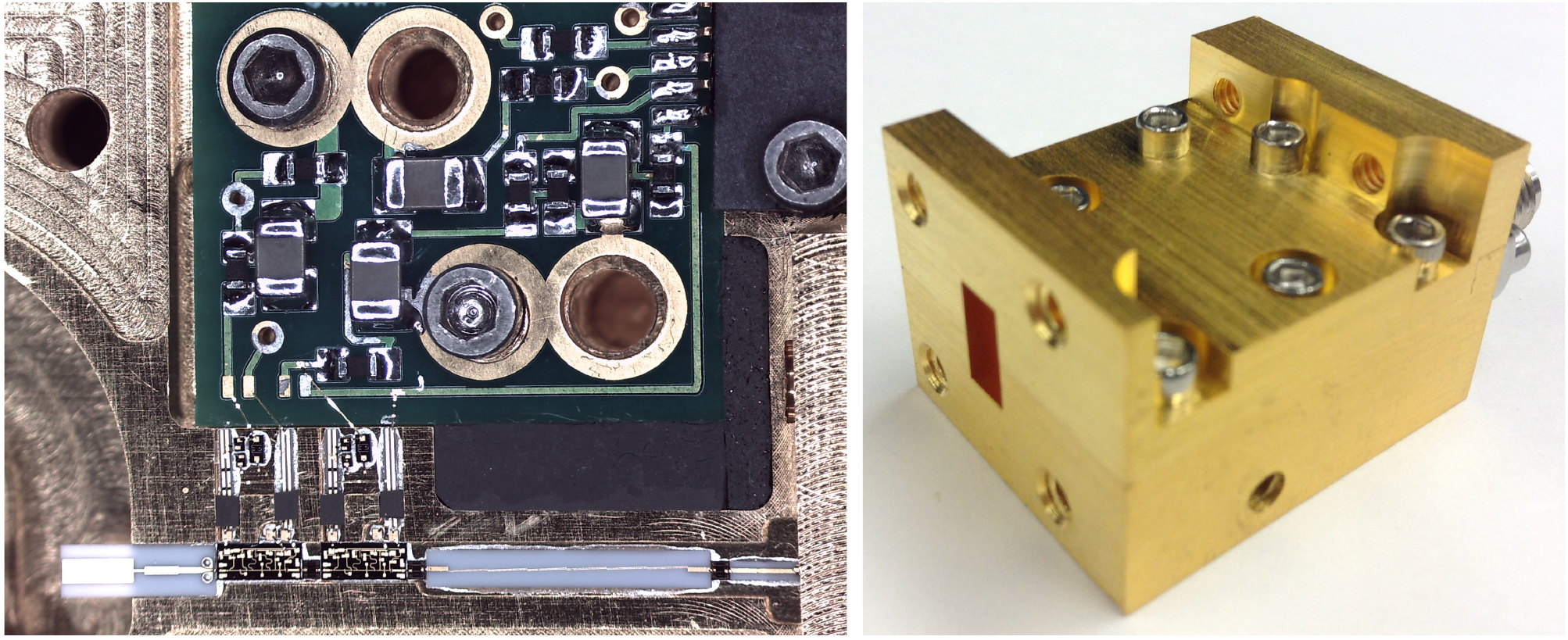}
    \caption{(\textit{Left}) Interior of the COMAP LNA module showing the chain of RF components and the bias conditioning PCB. (\textit{Right}) Module exterior showing the waveguide input.}
    \label{fig:lna_module}
\end{figure}

The module input is rectangular WR-28 waveguide in order to mate directly with the preceding rectangular-to-circular waveguide transition, as described above. Inside the module (see Figure\,\ref{fig:lna_module}), a waveguide probe couples the incoming RF signal to microstrip and from there it enters a chain of two MMIC amplifier chips isolated from each other with a 3\,dB attenuation pad. The amplifiers are followed by a high-pass filter designed to cut off the first local-oscillator frequency of 24\,GHz and prevent feedback of any such LO signal that leaks back into the front end. The output of the filter is then coupled to a coaxial bulkhead feed-through. Bias for the amplifiers is conditioned through a printed circuit board inside the module. Thermally isolating cables conduct the signals to the cryostat wall (\S\,\ref{subsubsec:cryostat}).

\begin{figure}
    \centering
    \includegraphics[width=1.0\linewidth]{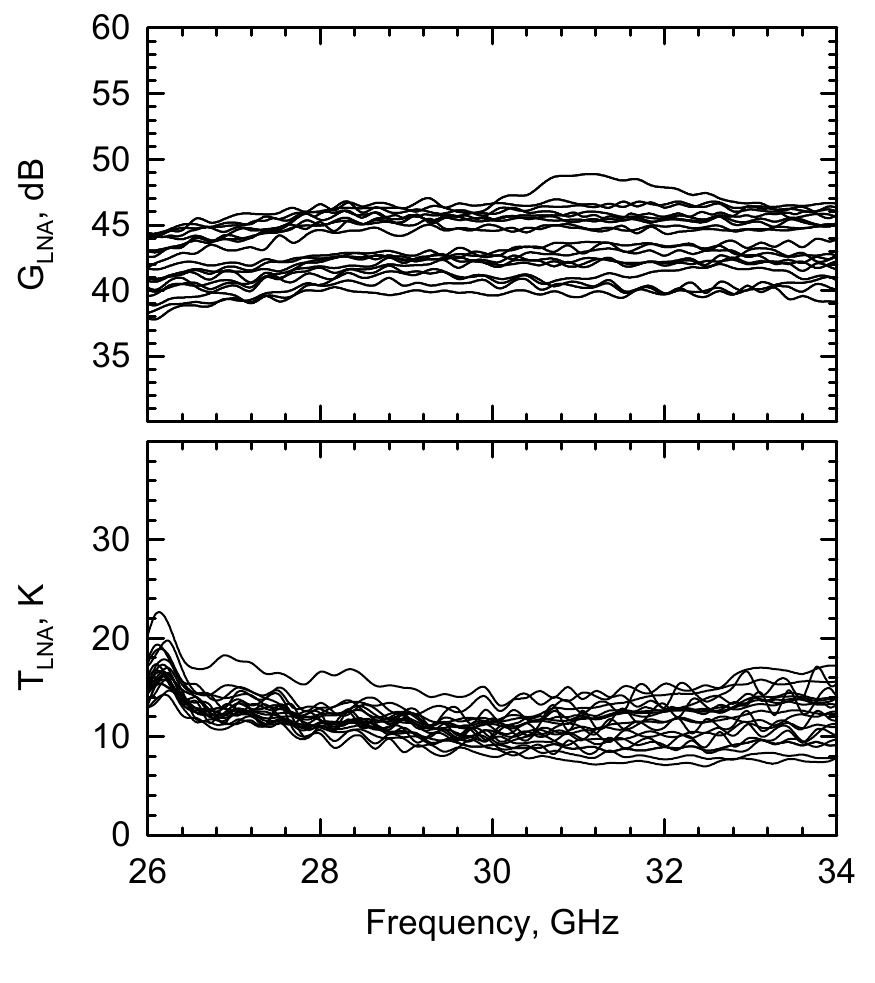}
    \caption{Gain (top) and noise temperatures (bottom) of a representative group of 19 LNA modules.}
    \label{fig:lna_module_perf}
\end{figure}

Each LNA module was characterized at cryogenic temperatures; Figure\,\ref{fig:lna_module_perf} shows the resulting gain and noise temperatures for the 19 modules used in the Pathfinder. The median values over the band ranged from 39.9--46.4\,dB and 8.1--14.4\,K, respectively, meeting the system requirements.

\subsubsection{First Downconverter (DCM1)} \label{subsubsec:dcm1}

Four shielded die-cast enclosures are mounted on the sides of the cryostat, each housing five downconverters. A 24\,GHz dielectric resonator oscillator (DRO) (Luff Research, PLDRO-24000-10) in each enclosure is phase locked to the 10 MHz source and drives five Marki ML1-1644IS mixers, and RF Lambda RLNA02G10GA amplifiers boost the signal for transmission via 15\,m cables to the sidecab. Heatsinks with fans on the outside of the modules conduct heat from the amplifiers. Each module has a control interface board that allows the amplifiers to be turned on or off, and to monitor the temperatures of the amplifiers.

\subsection{Sidecab} \label{subsec:sidecab}

Originally the sidecab accommodated millimeter receivers and the antenna control computer, and the signals were sent back to a central building for processing. For COMAP it was unclear what effects would be caused by a long optical fiber transmission to a remote building. Initial tests revealed some instability, but that was likely due to cycling of the air conditioning in the building. In any event, the decision was made to put the digital electronics in the antenna sidecab. The millimeter receivers were removed to allow for an elevation cable wrap, but the antenna control rack was left in place. The original thermal control comprising an air conditioner and a heater controlled by a commercial regulator and a blower keeps the sidecab temperature steady at 21\degr C.

Due to the volume of electronics, a ``pop-out'' extension was added (Figure\, \ref{fig:antenna}). An RFI-tight cabinet in the lower section contains the digital systems, and a split system air conditioner was installed in it to handle the additional heat load. The compressor located on the antenna platform has a variable-speed drive that maintains a very constant temperature. A small modification was required to over-ride one of its modes that causes significant temperature cycling during winter months.

DCM2 modules are housed in a card cage on top of the rack, with a fan to blow the temperature regulated sidecab air over them. LMR-240 cables connect the DCM2 outputs to a panel on top of the digital, and then to the ROACH-2 spectrometers.

\subsection{Cable Management} \label{subsec:cablemgmt}

Fifteen-meter-long LMR-400 cables run from the DCM1 modules on the cryostat exterior to the DCM2 modules in the sidecab. This choice was based on having a cable with acceptable loss variation across the 2--12\,GHz IF band, and a compromise between stiffness and flexibility to accommodate the elevation cable wrap without being too floppy. As mentioned above, the cables are routed from the receiver to the sidecab via an insulated duct. Once inside the sidecab, they are individually clamped to one straight edge of a plate curved with a 380\,mm radius. Passing over the plate, they hang off the other end in loops that attach to the sidecab wall. As the antenna changes elevation, the curved plate rotates with the tipping structure and the hanging loops accommodate the change in elevation angle.

From the clamps on the sidecab, the cables are carried by a cable tray over to the DCM2 modules at the other end of the sidecab.

Although ripple from this length of cable corresponds to $k$ values outside the range of interest, combining the 2\,MHz samples into 31.25,\ bins aliases this ripple from a period of 8.5\,MHz ($k_\parallel \approx 5.5$ Mpc$^{-1}$) to 100\,MHz ($k_\parallel \approx$ 0.5\,Mpc$^{-1}$). While this is in the relevant range,binning reduces the amplitude. Additionally, the cable loss and the 3\,dB matching pads at either end keep the ripple well below 1\%, and with the flex and temperature controls make it a negligible effect.

\subsection{Second Downconverter (DCM2)} \label{subsec:dcm2}

For each receiver, two downconverter boards are mounted on an aluminum plate for the two bands, A and B. The cables from the DCM1s go to SMA connectors on the front of the plate, generally with a coaxial attenuator selected to ensure that the following IF amplifier (RF Lambda RLNA02G10GA) is in its linear range. After the amplifier, a 16\,dB coupler allows a calibration tone to be injected, and then the signal is split in two, going to band-defining filters for bands A and B (Figure\, \ref{fig:signal}). Identical downconverter boards are used for the two bands.

Each board has a Marki MLIQ-0218SM IQ mixer followed by gain stages and digital attenuators (0.5 dB steps, 31 dB range), and log detectors for both I and Q. Two 0--2\,GHz low-pass filters are included to reduce aliasing and attenuate LO feedthrough. At the output a 3\,dB pad is used to improve the match to the ADC circuit in the digital processor which is connected with LMR-240 cable.

Local oscillators are either a 4\,GHz DRO (Luff Research PLOA4000-10), or an 8\,GHz DRO (Luff Research PLOA8000-10). Each is phase locked to the station 10\,MHz reference clock, and their outputs are distributed in a series of stripline power splitters.

The critical component for linearity in the system is the IF amplifier on the DCM2 plate. For each receiver, a pad between the cable and the DCM2 plate was chosen to ensure that the operating levels were in the linear range at the 1\% level when the ambient load was in place. Because of the variation in power from the receivers, there is a spread of about 10\,dB in pad values.

Cables from the DCM2 to the digital rack, and then from the rack to the ROACH-2s have total lengths up to 2.6\,m. For a typical cable temperature sensitivity of $5 \times 10^{-6}$ K$^{-1}$ Equation \ref{eq:rateofchange1} requires  stability on the order of 3\,K\,s$^{-1}$ at 2\,GHz, assuming a 100\,ms integration time and scan time of 60\,s. This is easily achieved.
 
\subsection{Digital Backend} \label{subsec:dbe}

Thirty-eight ROACH-2 systems process bands A and B from the 19 on-sky pixels. Some modifications were made to the chassis so that all connections and buttons, including power, are on the same end of the chassis, and a splitter was installed to feed both I and Q channels with the sampling clock. These chassis were mounted vertically with 13 on each of three shelves in the digital rack. Sheets of conductive foam between the chassis cushion them from telescope vibrations and provide additional RFI protection.

IF levels going into the ROACHes are set to give an rms count at the 8\,bit ADC (e2v semiconductors, EV8AQ160) of $\sim$~42 with the calibration vane inserted into the beam. On the sky, the power drops by almost an order of magnitude so that the rms is $\sim$~14, allowing a significant margin above the digitization noise level. A Valon 5009 synthesizer generates the 2\,GHz sampling clock and the ADCs trigger on rising and falling edges to sample the input at 4 GHz.

Data collection is initiated by sending a signal to the ROACHes to start integrating on the next hardware 1\,PPS (pulse per second) signal. 8192 samples are then captured from each ADC and transformed to the frequency domain by a poly-phase filter bank (PFB) \citep{price_spectrometers_2021} to produce 1024 complex frequency channels spaced by 1.953125\,MHz.

Weighted combinations of the complex I and Q values yield the lower and upper sidebands. After squaring the channel amplitudes, 39063 power spectra are accumulated for for a 20.000256\,ms integration time. Consecutive integrations follow on without any dead time. Final requantization of each integration reduces the value to a 24\,bit integer without risk of overflow, and the LSB and USB spectra are transmitted with a sequence number to the host computer through a 10\,Gb switch using UDP.

\subsubsection{Polyphase Filter Bank Implementation} \label{subsubsec:pfb}
\begin{figure}
    \plotone{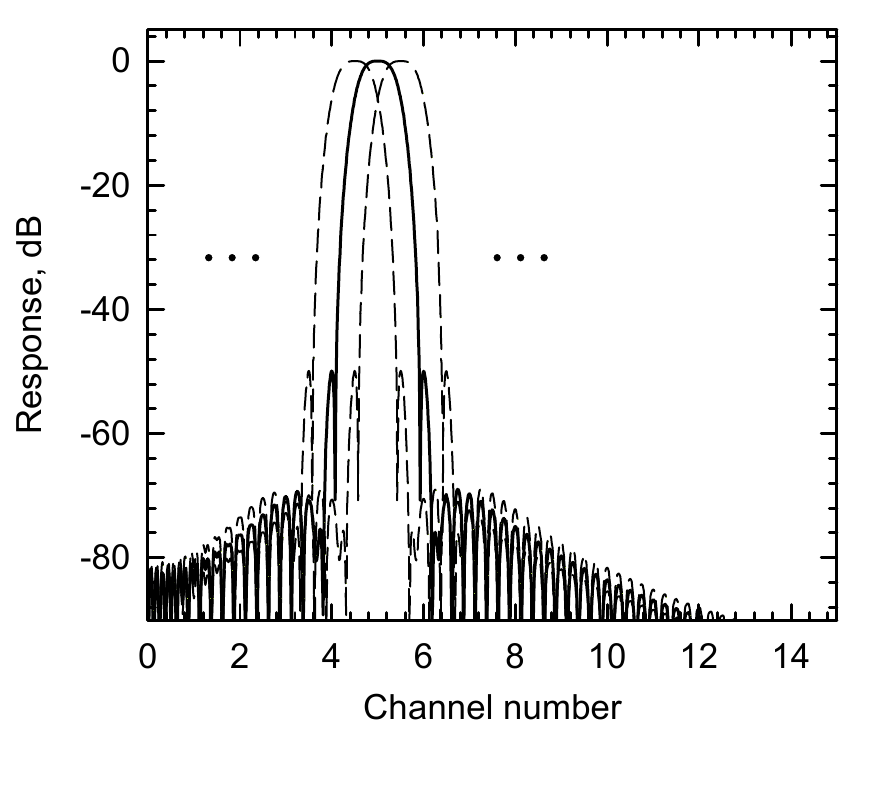}
    \caption{Effective channel passbands from the poly-phase filter bank.}
    \label{fig:specpassband}
\end{figure}

The four-tap PFB (see, for example, \citealt{price_spectrometers_2021}) takes 8192 voltage samples and applies weights given by the product of a Hamming window and a sinc function:
\begin{equation}\label{eq:weights}\begin{split}
    w_n &= \left (0.46164 \cos \left ( 2 \pi \frac {n - \frac{N + 1} {2}} {N} \right ) + 0.53836 \right ) \\
    &\quad\cdot \sinc \left (4 \pi \frac{n - \frac {N+1} {2}} {N} \right).
    \end{split}
\end{equation}
Figure\,\ref{fig:specpassband} shows the resulting passband.Although the standard calculation of the noise equivalent bandwidth indicates a sensitivity loss of $\sim$\,10\%, simulations predict a value of only 2.5\%. This is perhaps related to the nature of the PFB, in which consecutive spectra overlap by 75\%.

\subsubsection{IQ Separation and Calibration} \label{subsubsec:iqcal}
To achieve a high degree of fidelity in the LSB and USB spectra a set of \textit{IQ coefficients} is used to combine the I and Q signals \citep{finger_calibrated_2013}. Initially, these coefficients are set so that there is no separation of I and Q spectra. A tone from a Hittite HMC-2220 frequency synthesizer is injected at the center of each frequency channel and the amplitudes and phase difference of I and Q are measured. These are used to compute the IQ coefficients which are then written to the FPGA. Thereafter, when the coefficients are used to weight the sums of I and Q the resultants are the sideband-separated LSB and USB spectra.

As a practical detail in the calibration, the DCM1 IF amplifiers are turned off to remove most of the noise component. Injection of the test tone at a power level comparable to the normal noise power level exercises a similar range of bits in the ADC, but all the power is concentrated in a single channel of the PFB. With the normal scaling this would overflow the corresponding register so a different rescaling is implemented relative to the normal scaling for noise.

\subsubsection{Aliasing} \label{subsubsec:aliasing}

\begin{figure}
    \plotone{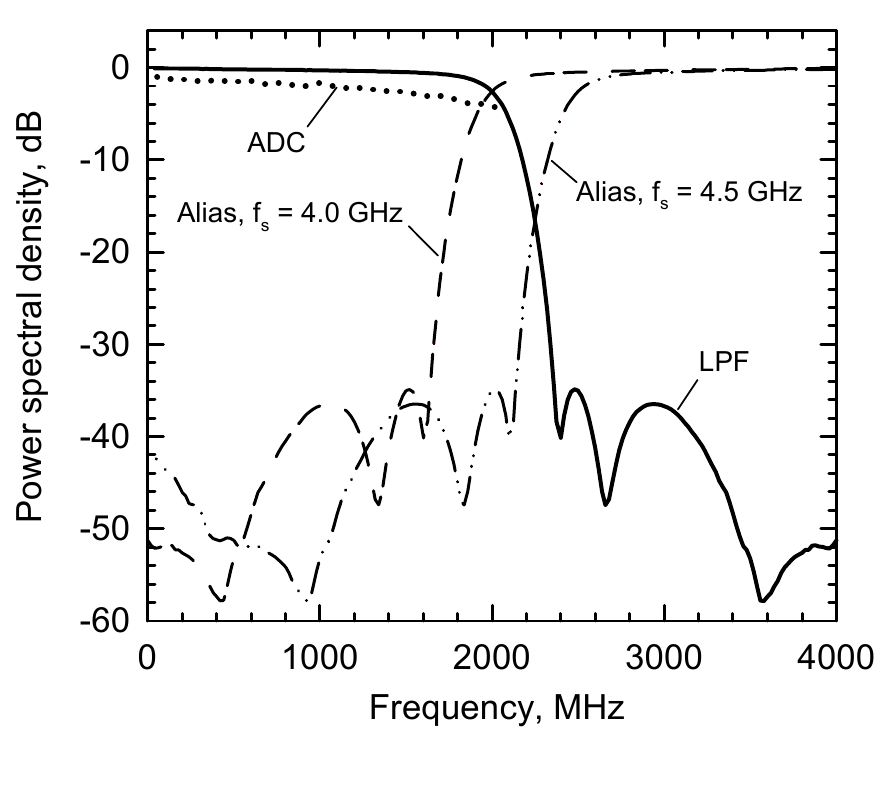}
    \caption{A Mini-Circuits low pass filter (LPF), model LFCN-1700, is used to attenuate signals potentially aliased into the 0--2\,GHz base band into the EV8AQ160 ADC. The ADC response falls off slowly according to its data sheet. Currently, the sampling frequency folds in significant aliased power, but a proposed increase to 4.5\,GHz will essentially eliminate this.}
    \label{fig:antialiasing}
\end{figure}

Aliasing of signals outside the nominal spectrum occurs because the band-defining filters for bands A and B, and the low-pass filters in DCM2 do not efficiently filter out signal close to the band edge, as illustrated in \ref{fig:antialiasing}. In part this is because filters with very sharp cutoff were avoided since the high number of poles required makes them more temperature sensitive, potentially leading to subtle changes in the passband shape. Currently, channels where the contribution of unwanted signal is more than $\sim$~$-15$\,dB relative to the in-band signal are discarded. A better solution is to sample at 4.5\, or even 4.25\,GHz so that the aliased power is attenuated more by the filters, with most of the aliasing being folded to frequencies outside the band of interest. It is a significant challenge to meet the timing in the FPGA so that was deferred until resources were available. That work is currently underway.

\vspace{8mm}
\subsection{Monitor and control} \label{subsec:mnc}

Much of the antenna-based monitor control system electronics were taken directly from CARMA \citep{woody_carma:_2004}, with some additional components based on a control system for Argus \citep{sieth_argus_2014}. A VME crate with a CPU running Linux (CentOS 6.5,  32\,bit) acts as a master in a controller-area-network bus (CANbus) \citep{marsh_canbus_2002} distributed system, with a protocol developed for CARMA \citep{woody_controller-area-network_2007}. CANbus modules include encoder and drive interfaces, environmental monitoring, calibration vane interface, cryogenic compressor monitoring, etc. Other parts of the system, such as the calibration signal synthesizer and the ROACH-2 modules communicate directly with the central control system through the antenna 10 GbE switch.

Following the CARMA scheme, most monitor points are in a 0.5 s cadence ``blanking frame''. COMAP places an additional requirement on the monitoring because the rapid scanning needs a faster update of the antenna pointing. In its fast motion modes, the antenna is unable to precisely track the commanded positions, but the actual position is always known accurately. Encoder readings are strobed on every 100\,ms boundary (see Section \ref{subsec:time} for details of timing) and are reported in sets of five per the 0.5 s packet. Simple interpolation on to the 20 ms integration interval is more than accurate enough for mapping the data to position on the sky.

\vspace{5mm}
\subsubsection{Warm Support Electronics} \label{subsubsec:supelec}

Custom warm support electronics control and monitor the focal plane array’s frontend receiver package with 38 independently settable bias voltages for the LNAs and monitors. A second system controls the 38 DCM2 second-stage downconverters, including their environmental monitor points and their LO phase-lock states. Two Netburner Mod5270 microcontrollers, one for the frontend and one for the backend, communicate with the COMAP control system by Ethernet, and with the electronics through I2C serial commands. To suppress RFI, both microcontrollers are in individual shielded enclosures with media converters for fiber Ethernet connections and RF filtering on the 100\,kHz I2C signal lines.  Controlling individual components with I2C internal interfaces through sub-buses selected by cascaded I2C switches has proved to be a compact and practical solution to reaching the several hundred control and monitor points in the system. 

We took several steps to reduce pickup and other noise in the frontend LNA bias system, minimizing the effects of connections to the bias electronics on overall system noise. Circuit cards for bias, I2C bus switches, LNA power control, and cryogenic diode temperature current sources and readouts are on a simple backplane housed in a small card cage mounted on the cryostat, connected by $\sim$~1\,m-long cables to the cryostat and by a $\sim$~10\,m-long cable to the control microcomputer in the receiver cabin. A 1:44.5 resistive voltage divider between the bias system and LNA gates attenuates pickup from outside the cryostat and limits the gate voltage to 0.3\,V even if the bias voltage swings to an op-amp rail. Series diodes clamp the drain voltages to a safe 1.4\,V, and the drain bias circuit limits the output current to 50\,mA. Soft limits in the microcontroller firmware further ensure that the bias voltages on the gates and drains are within safe limits for the LNAs. Mechanical relays under microcontroller control can electrically isolate the bias electronics and LNAs, and allow power sequencing under microcontroller control. 

Additional electronics in the frontend system control and monitor the power supplies, sources 10\,\textmu A current for the cryostat’s temperature-sensing diodes, connects with the cryostat’s pressure gauge, and measures temperatures within the electronics. Small cards in each DCM1 module control and monitor IF amplifier power and temperatures. The backend system provides power and communication for each of the 38 DCM2 modules, and also monitors the phase-lock bits for the downconverter LOs.    

\vspace{3mm}
\subsubsection{Control Software} \label{subsubsec:antctrl}

The antenna control system is an amalgam of two pre-existing systems. Code from the CARMA software system \citep{scott_carma_2004} that ran in the antenna Linux computer was modified to communicate with a user interface and archiving system based on code used in several other telescopes at OVRO and elsewhere \citep{padin_cosmic_2002,story_south_2012}. The observer is presented with a flexible user interface that has a command window and user-definable monitoring windows, with either numerical fields or time series plots. For the most part, operations are run through a built-in scheduling language that can also invoke Python scripts for easy development of functionality.

Monitor points from CAN modules in the antenna are pushed out every half-second and collated in the CARMA-based antenna code. These are filled into registers in the central control system where they are recorded in a flat file system. Additional housekeeping data, such as pointing model corrected azimuth and elevation, source name, commanded position, are combined with the CANbus monitor points.

The control system also interrogates the warm support electronics for monitor data from the frontend and the DCM2 modules on a 15\,s cadence, and data from the site weather station are included in the registers.

Much of the antenna control code was taken directly from the CARMA system. Inputs from the main control system are at a high level since the antenna code can take input in the form of RA and Dec, azimuth and elevation, and offset pointing in various coordinate systems. Conversion from J2000 coordinates to antenna azimuth and elevation, including refraction correction is performed by the antenna computer. It also applies the pointing model, described below.

Leighton's original servo loop \citep{leighton_final_1977} is used to close the position loop. It takes the current and predicted positions at consecutive half-second intervals and derives velocities to apply to the azimuth and elevation motors. Proportional, integral, and differential terms are used to improve tracking accuracy.

Some special modes were implemented for COMAP, allowing the antenna to do horizontal raster scans or Lissajous figures on the sky. In these cases, the tolerance on the acquisition of the commanded position is relaxed and the telescope does not closely follow the nominal trajectory, but the position is recorded accurately every 100\,ms.  Raster scans are implemented by acquiring the source and then alternating positive and negative offsets, allowing the antenna to move between them at a speed determined by its natural acceleration and velocity limits. The Lissajous mode, where sinusoids with individual amplitudes, frequencies, and phases are applied to the azimuth and elevation axes, are less stressful than the raster scans, as well as giving better field coverage near transit. The relative merits for the data quality are analyzed in \citep{es_III}.

\subsection{Time and Frequency References} \label{subsec:time}

UTC (Universal Coordinated Time) is derived from the OVRO Meinberg Lantime stratum 1 GPS time server. Computers on the OVRO network synchronize to this using NTP (Network Time Protocol), typically maintaining time to better than a millisecond. A CARMA master clock CAN module takes the 10\,MHz and 1\,PPS (pulse per second) outputs from the GPS clock and uses them to discipline its Rb clock (SRS PRS-10) with a four hour time constant, giving excellent short- and long-term stability. All systems requiring 10\,MHz and 1\,PPS inputs use the output from the Rb clock. Components that require the 10\,MHz signal include the sampling clock for the digital data acquisition, the sampling clock frequency counter monitor, the first and second downconverter oscillators, and the Hittite calibration signal synthesizer.

Precise phase-locking of all oscillators is critical to the system stability. Small changes in LO frequencies shift ripple slightly producing residual ripple in the data.

Matching the time between the encoder readings and the data integrations is the most critical aspect of timing. All of the CAN modules in the system keep Modified Julian Date (MJD) time internally. Time stamps are sent out on the CANbus from the Linux host computer approximately every ten seconds. The exact time is read from the NTP clock when the time message is transmitted, and since the time stamp messages have the highest priority on the bus, there is negligible latency, ensuring that timing is accurate to about a millisecond. Internally, the CAN module servos its timer to keep its clock synchronized without introducing discontinuities. Encoder readings are triggered by an interrupt from its MJD clock every 100\,ms, and after five readings are acquired, they are sent back in the blanking frame packet on the 0.5\,s boundary. The time assigned to each encoder reading is its nominal exact 100\,ms time boundary, so latency in returning packets does not affect the accuracy. Note that this assumes timing on the host is tight enough to avoid ambiguities in identifying the correct 0.5\,s frame, which has been verified over a decade of operation in CARMA.

Accurate timing for the data integrations relies on the 1\,PPS from the master clock, which is fed to each ROACH-2 chassis. When data acquisition is initiated from the control system, a ``dada'' ring buffer \citep{van_straten_psrdada_2013} is set up to capture UDP data packets on the 10 GbE network. The header is set up, and in the half second before the designated start time a command is sent to the ROACH-2 PPCs to arm them to start acquisition on the next 1\,PPS rising clock edge. Although the ROACH-2s are not guaranteed to start on the same 4\,GHz sampling clock cycle, temporal offsets much smaller than the integration time have negligible effects.

Verification of the precise alignment of the CAN module time with the PPS was done by temporarily injecting a pulse generated by the CAN module software into one of the feed ROACH-2 ADC inputs and verifying that it appeared at the correct time in the data stream. This has a modulo one second ambiguity, which does not detect the case where a ROACH-2 misses its trigger time and is delayed by a second. That case can be detected by comparing the times that the spectrometer total power changes for all feeds as the calibration vane is inserted and removed. Affected data flagged and removed from the analysis; this is infrequent enough that it is not deemed useful to try to realign the data after the fact.

\subsubsection{Pointing} \label{subsubsec:pointing}

Pointing of the 10.4\,m telescope has been extremely well characterized over decades of operation. An optical telescope is permanently mounted on a representative part of the backing structure and observes through a hole in the primary. A series of stars is observed over the sky, with the observer centering each in the field of view and recording the necessary offsets. An eight parameter model is fit to these data. Five of these refer to the mount and are common to the radio pointing. Three (elevation, cross-elevation collimation and sag) must be separately measured for the radio beam through radio pointing. Sag is the most difficult term to determine for radio pointing but once determined from a dedicated run over a large elevation range, it is stable enough not to require recalibration.

Tilting of the antenna azimuth axis driven by solar heating is significant, but a tiltmeter located on-axis continuously monitors the two orthogonal tilts and the values are applied to the pointing model in real time. At intervals of a week or so, the telescope is rotated through 360\degr \ in azimuth and the co-rotating tiltmeter readings analyzed to derive a constant offset between the tiltmeter axis and the azimuth axis. This is particularly important before a pointing run.

In general, blind pointing is accurate to $\leq$\,0.2\,arcmin, and offset pointing to $\leq$\,0.02\,arcmin, more than sufficient for COMAP science.

\subsection{Data Acquisition and Processing}

When data acquisition is started by the control system a software package, PSRDADA \citep{van_straten_psrdada_2013}, is initialized with header information including such details as the start time, integration time, and other necessary metadata. Ring buffers are initialized to capture spectrometer data pushed out by the ROACH-2s as 24 bit integers, and they are combined into 20\,ms frames. If packets are not captured within a specified time, the data are filled in with a special value. Data are streamed into a ``dada'' file. Typically, observations are split into one hour chunks to create manageable size data files.

Once a dada file is complete, it is sent to campus to process into a ``Level 1'' file. This conversion includes: 
  transposing the ordering of feed and time;
  converting spectra from 24\,bit integer to 32\,bit floats;
  replacing ``special'' 24\,bit numbers (e.g., overflow) to ``NaN'';
  including monitor point archive values;
  calculation of RA and Dec from azimuth and elevation for each feed;
  calculation of system temperatures.

The monitor point archive includes ``feature bits'' that are set by the control system to indicate particular observing modes. For example, they may indicate whether raster scanning or Lissajous scanning is in effect, or if the measurement is a sky dip or a system temperature measurement. The Level 1 conversion uses these bits to group or process data in particular ways, and they are also available for later stages of processing.

\section{Performance} \label{sec:perf}
\subsection{RF Performance} \label{rfperf}

System noise performance was evaluated by scanning the telescope in elevation on the sky, bracketed by an ambient load measurement at the start and end. For each 20\,ms integration, the system temperature and gain were determined for every 2\,MHz channel using the ambient vane calibration technique. This system equivalent noise temperature, which is referred to outside the atmosphere, can be written as
\begin{align} \label{eq:tsys}
    T_\mathrm{sys}= &\left(T_\mathrm{rx} + \left(1-\eta_\mathrm{sp} \right) T_\mathrm{amb} + \right.\nonumber\\
                        &\left.\eta_\mathrm{sp}\left(1-\mathrm{e}^{-A \tau_\mathrm{z}} \right) T_\mathrm{atm}\right)\eta_\mathrm{sp}^{-1}\mathrm{e}^{A \tau_\mathrm{z}} + T_\mathrm{cmb},
\end{align}
where $T_\mathrm{rx}$ is the receiver noise temperature, $\eta_\mathrm{sp}$ is efficiency due to spillover terminated at ambient temperature ($T_\mathrm{amb}$), $\tau_\mathrm{z}$ is the zenith opacity, $A = 1/\sin{\theta_\mathrm{el}}$ is the number of air masses as a function of the elevation angle ($\theta_\mathrm{el}$), $T_\mathrm{atm}$ is the physical temperature of the atmosphere, and $T_\mathrm{cmb}$ is the brightness temperature of the cosmic microwave background.

\begin{figure}
    \includegraphics[width=0.475\textwidth]{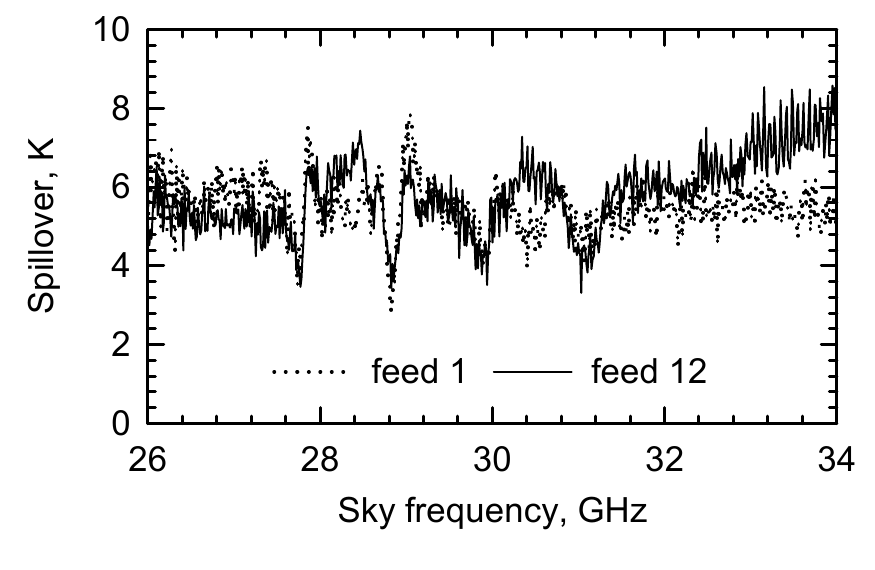}
    \caption{Zenith spillover noise estimated by taking the difference in system temperature with the secondary support absent and installed.}
    \label{fig:spill}
\end{figure}

By fitting the data to Equation~\ref{eq:tsys}, the system parameters can be determined. However, there is sufficient degeneracy among some of the parameters that we fix $T_\mathrm{atm}=T_\mathrm{amb} - 20$\,K, and $\eta_\mathrm{sp}=97.5\%$. The spillover is consistent with a measurement made when the secondary support was removed for modification. System temperatures measured with and without the secondary assembly in place yielded an estimate of the spillover noise at zenith. For the two feeds installed in the initial configuration, the spillover was found to be $\sim$~5--6\,K (Figure\,\ref{fig:spill})
In reality, the spillover is a function of elevation, so it is partially degenerate with the atmospheric elevation dependence. For now we assume for simplicity that the spillover is a fixed number.

\begin{figure*}[ht!]
    \includegraphics[width=1.0\textwidth]{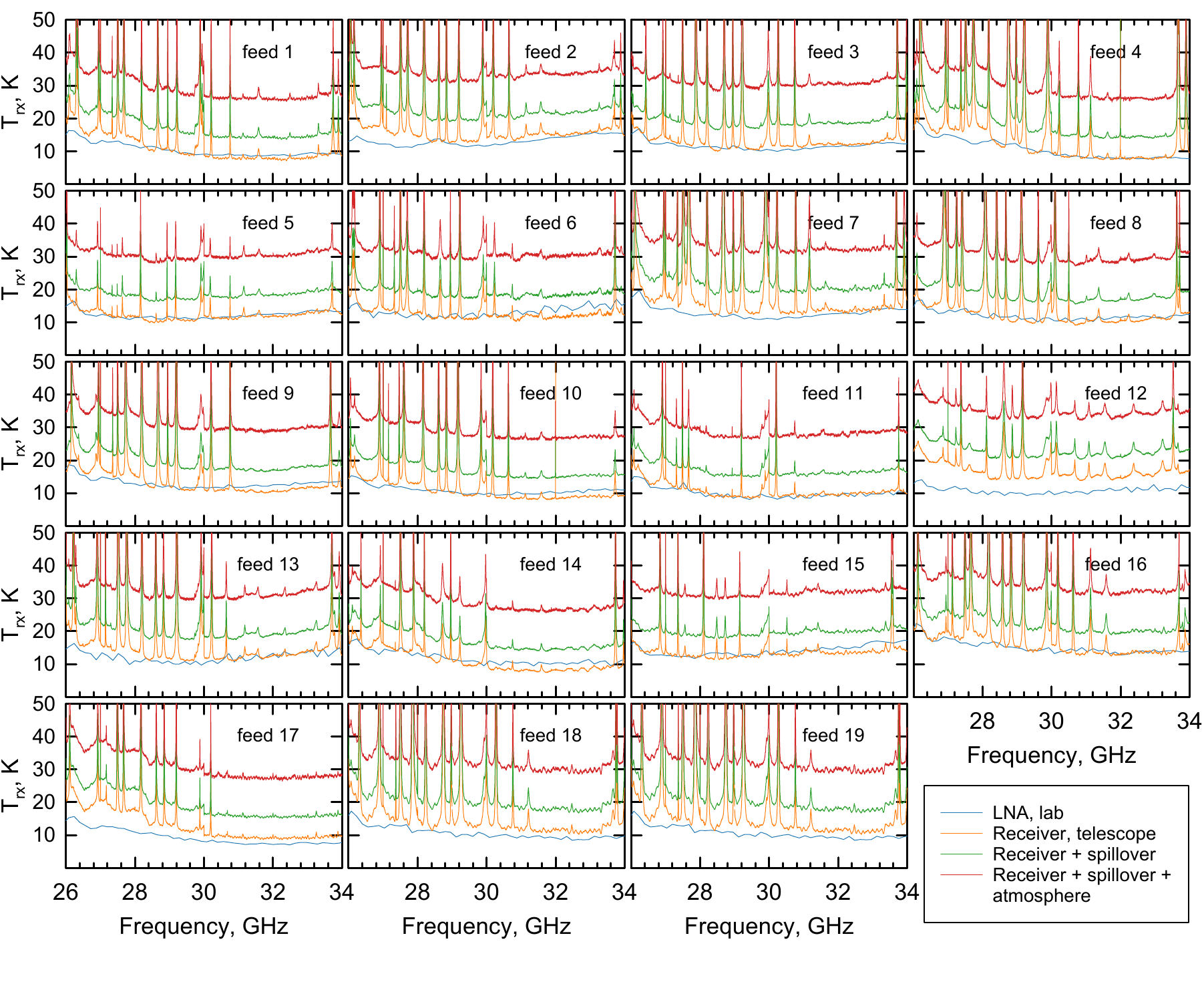}
    \caption{System noise temperatures. The blue curves are the LNA noise temperatures measure in the lab. The orange curves show the values derived from the sky dip, while the green curves also include the estimated spillover to ambient. Curves in red include all noise components including the zenith atmospheric brightness component. See text for discussion of the noise spikes.}
    \label{fig:trx}
\end{figure*}

Noise temperatures derived for all the receivers channels are shown in Figure\,\ref{fig:trx}. In general these are excellent, but an obvious issue is the very prominent noise spikes, particularly in the lower half of the band. The precise cause of these spikes is presently unclear, but it is known to be an interaction between the polarizers and the corrugated horns. When the polarizer is replaced by a plain circular waveguide, or the corrugated horn is replaced by a smooth-wall conical horn, the spikes disappear. Return loss measurements also show some evidence of the spikes, but electromagnetic simulations with HFSS with some tolerance errors introduced have so far failed to show similar behavior. An experiment with an attenuating septum in the circular waveguide between the rectangular to circular converter and the polarizer to terminate the orthogonal linear polarization resulting from reflections did not show any improvement. Similar phenomena have been reported by, for example, \citet{morgan_graphical_2013}, which will guide a resolution of the problem in future instruments.

In the interest of expediency, the receivers were installed in this condition, and the spikes are flagged out from the data. They are fairly stable in frequency and amplitude, but they have been observed to vary slightly with bias changes in the first stage of the LNA.

\begin{figure}
    \includegraphics[width=0.45\textwidth]{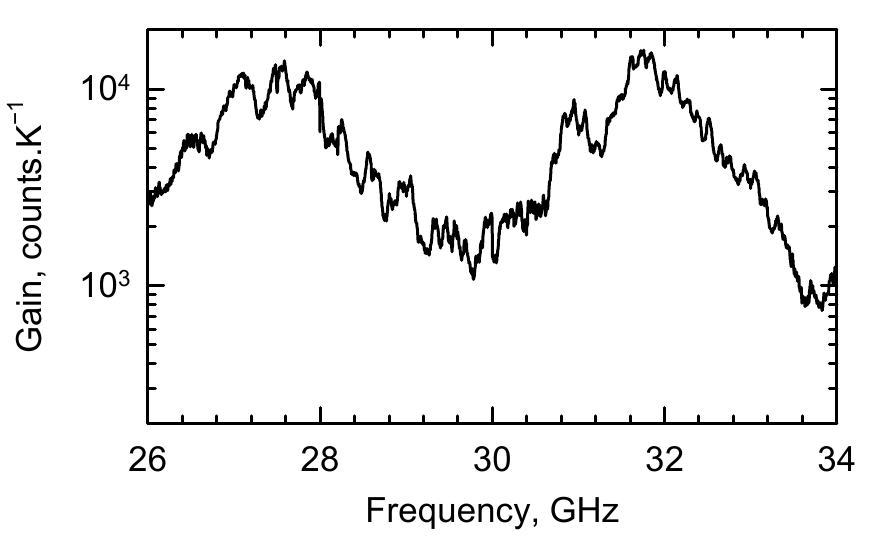}
    \caption{Median gain for the 19 receiver channels.}
    \label{fig:sysgain}
\end{figure}

Figure\,\ref{fig:sysgain} shows the gain of the receivers represented by the median over all the feeds. Spikes due to sampling artifacts have been removed, but there is a significant amount of structure remaining. Some of the uncertainty in the noise and gain measurements results from the calibration load variability. It was understood at the start that it would be almost impossible to build a calibrator to precisely determine the system passband

The intention was to use the scan data to remove structure in the passband and noise temperature and to use the calibration vane to obtain the overall scaling to brightness temperature units. Differencing ambient load measurements taken  one minute apart, at 40\degr \ and 60\degr \ elevation, revealed a ripple component corresponding to the LNA-load distance. The temperature was stable to better than the resolution of the temperature sensor, 0.2\,K, or 0.07\%, and the mean difference in power for the two measurements was between $-0.42$\% and 0.12\%, depending on the feed. However, the standard deviation of the ripple varied among feeds from 0.23\%--1.49\%. This has little impact on the CO line mapping results or the Galactic survey.

\begin{figure}
    \includegraphics[width=0.45\textwidth]{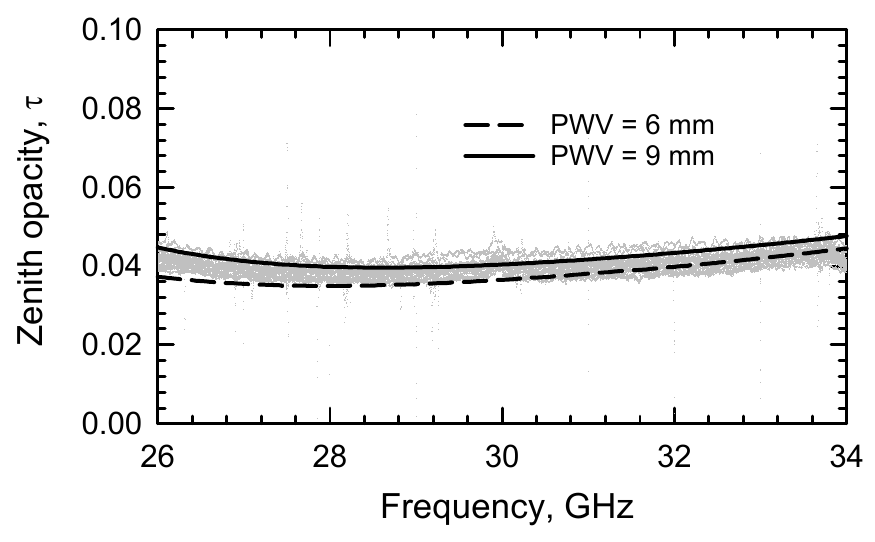}
    \caption{Atmospheric zenith opacity derived from an elevation scan on the sky on a good observing day.The ensemble of feeds is shown in grey, while the dashed and solid lines are modeled with PWV of 6 and 9 mm respectively.}
    \label{fig:tatm}
\end{figure}

\subsection{Spectrometer Performance} \label{subsec:dbeperf}

The analog signal chain is set up with the calibration vane inserted to give a noise input to the ADCs such that the $\pm 3\sigma$ levels are at the limits ($-128$ to $+127$ digitizer counts) so that $\sigma \approx 42$ counts. On the sky the power is about 10\,dB lower, so $\sigma \approx 14$ and the digitization noise is negligible. At this setting the digitizer is accurate on both the sky and the calibration load, and has a sufficient dynamic range to cope with the $\sim$~10\,dB power variation across the spectra.
\begin{figure}[hb!]
%\epsscale{1.15}
    \includegraphics[width=0.45\textwidth]{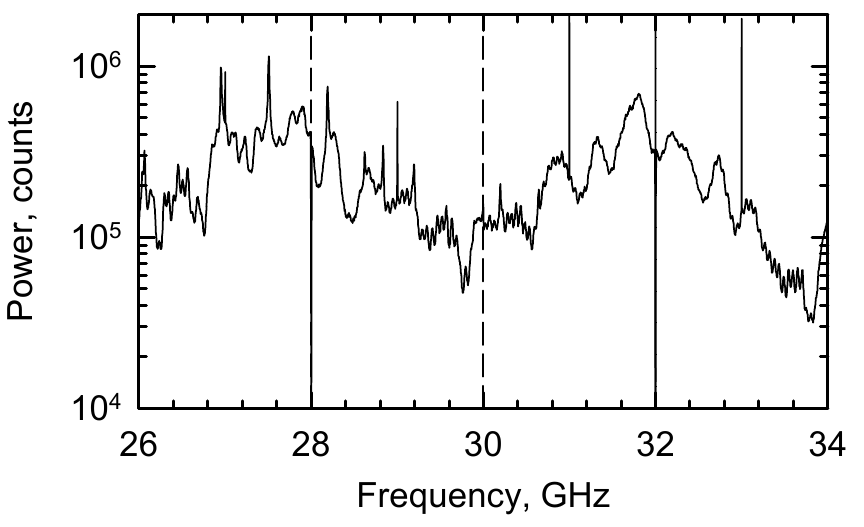}
    \caption{Example of a raw spectrum. Dashed lines indicate boundaries of the four sidebands}
    \label{fig:rawspec}
\end{figure}

A typical raw spectrum is shown in Figure\,\ref{fig:rawspec}. The end channels of each sideband, corresponding to DC and its Nyquist alias are discarded. The discontinuity at the center of the sidebands are artifacts of the four-core architecture of the ADCs, while the discontinuities at one quarter and three quarters of the channel range are attributed to the DC component of the 0--2\,GHz baseband signal, or the 6 and 8\,GHz LO signal at the IF. These frequencies are also discarded.

During observing, the IQ coefficients for separating the sidebands of the DCM2 LOs are periodically determined. This is essential when the ROACH-2s are power-cycled, but in general the coefficients are very stable over time. Power-cycling the ROACH-2s causes them to lose clock alignment and an arbitrary shift of up to 16 integer clock cycles can be introduced. In principle, only a coarse sampling of 16 frequencies is sufficient to determine this and the previous coefficients can be modified accordingly, but since this is required infrequently it has not been worth the time to implement.

\begin{figure}
\centering
\gridline{\fig{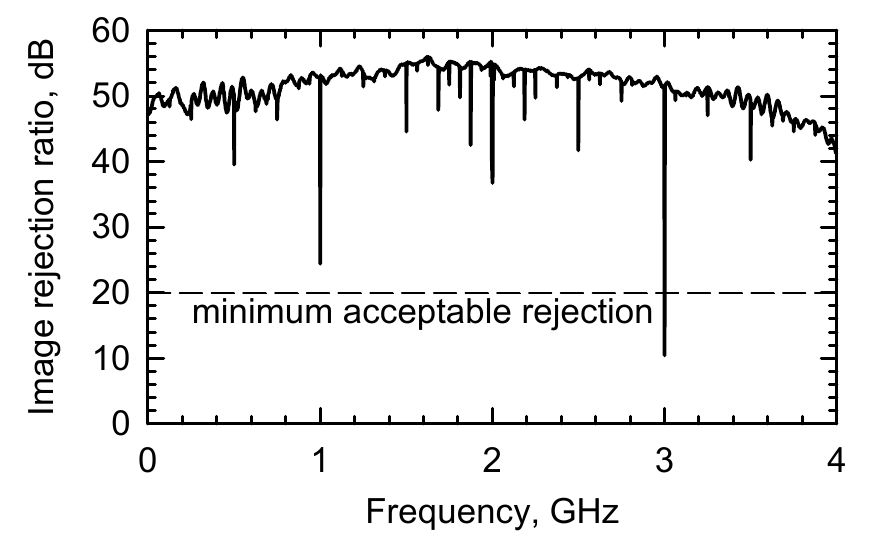}{0.45\textwidth}{(a)}}
\gridline{\fig{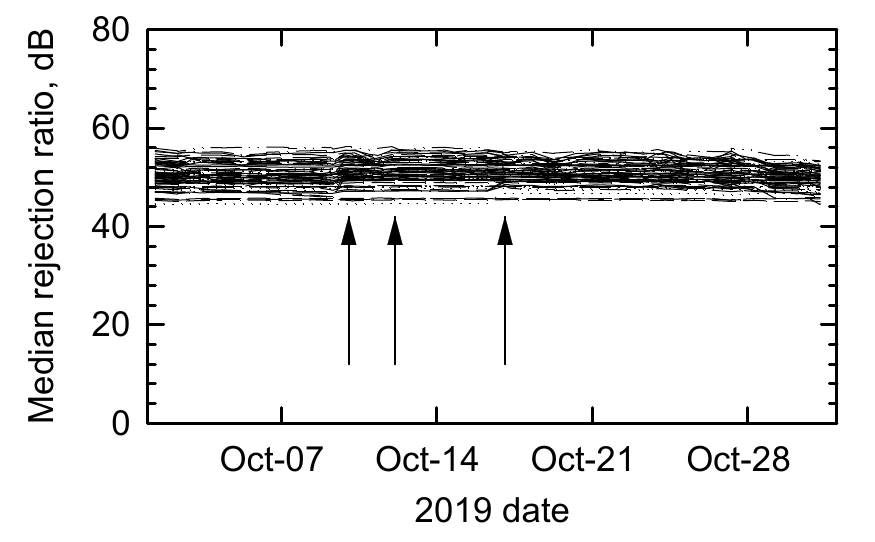}{0.45\textwidth}{(b)}}
\caption{(a) A typical plot of the image rejection ratio across the two 2\,GHz sidebands of a second IF. The requirement is 20\,dB, which is easily met except at frequencies that are rejected for other reasons. (b) The evolution of the median rejection ratio for all the bands in the system over a one month period. Arrows indicate time that the IQ coefficients were remeasured, usually because of a system restart.}
\label{fig:iqmeas}
\end{figure}

Figure\,\ref{fig:iqmeas} shows the achieved rejection ratio after calibration for a typical band. The 20\,dB specification is easily met at all frequencies of interest. Furthermore, the calibration is stable over the longest period that has been observed between calibrations (of order a month).

\subsubsection{Aliasing} \label{aliasingmeas}Aliasing was a known artifact of the system during the design phase. Sampling at 4\,GHz was intended as an interim solution, with a goal of 4.25\,GHz sampling (\S\,\ref{subsubsec:aliasing}). Tests showed that the current FPGA implementation does not support sampling more than a few percent higher than 4\,GHz, so the aliasing was quantified in detail using the injected test tone.

\begin{figure}
    \centering
    \includegraphics[width=0.425\textwidth]{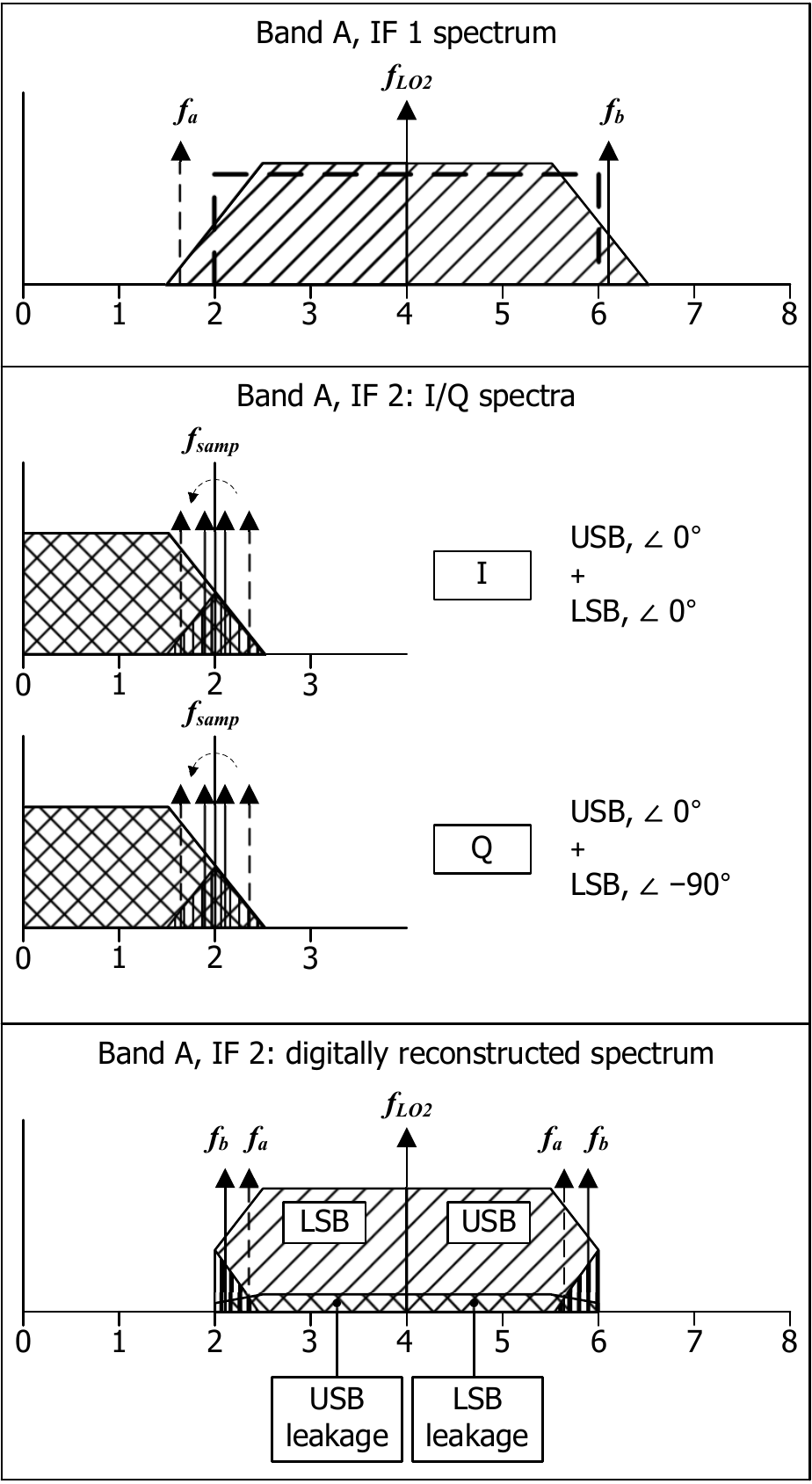}
    \caption{Diagram illustrating the effects of signal aliasing. The top panel shows the input from IF 1 to the second downconverter. The dashed box indicates the nominal band. In the second panel, the signal has been converted down to the 0--2\,GHz baseband (IF2), folding the USB and LSB bands together along with the aliased signals. Finally, in the bottom panel the digitally separated signals are shown with their assigned frequencies. Imperfect sideband separation is represented by the bottom hatched region, and the aliased signals are depicted by the arrows.}
    \label{fig:aliasingdiagram}
\end{figure}

\begin{figure}
    \centering
    \includegraphics{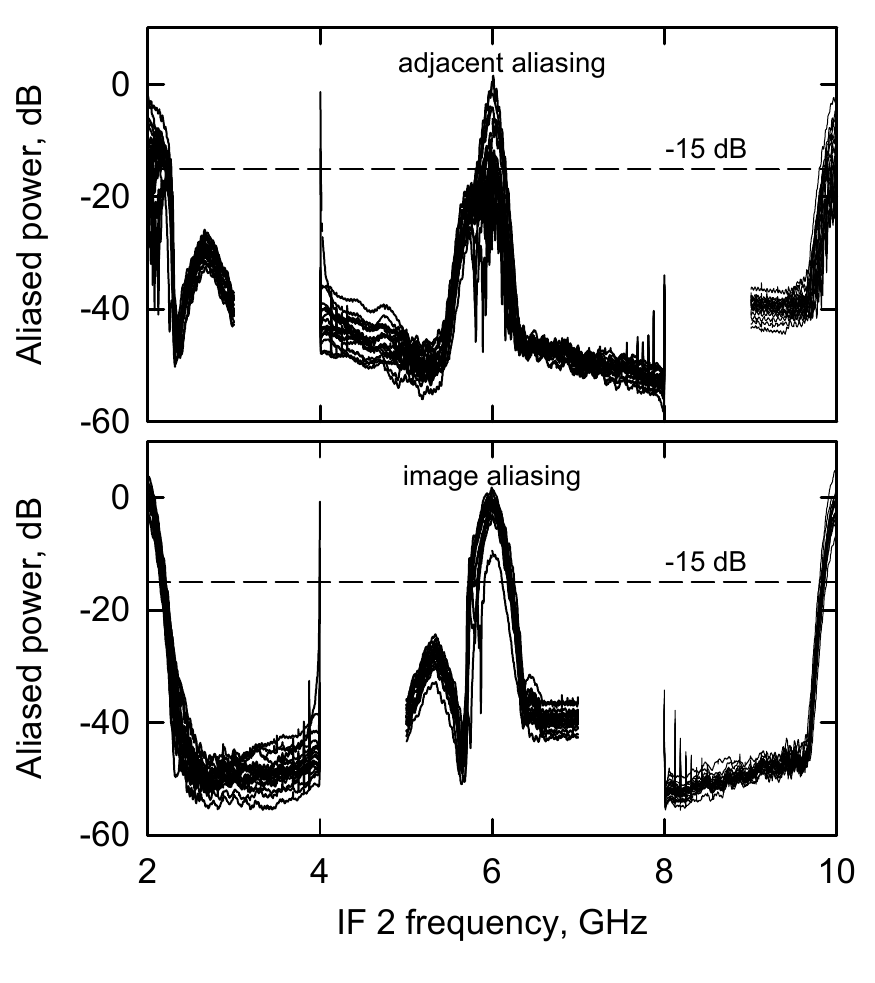}
    \caption{Measured aliasing using an injected tone. The top panel shows the leakage from signals in the same sideband as the nominal signal, while the lower panel shows signals leaking from the opposite sideband. The IQ coefficients applied are correct for the nominal signal, but should not be expected to be appropriate for the image, which is 4\,GHz away. The I and Q path lengths in DCM2 were not equalized, which would probably have given better cancellation of the aliased signal also.}
    \label{fig:aliasingresults}
\end{figure}

As expected, there is significant leakage from signals outside the band being reflected into the band. Contamination from just outside the opposite edge of the band was also found in the measured data (see Figure\,\ref{fig:aliasingresults}). This results from aliasing into the opposite sideband that is not rejected by the IQ separation procedure, as illustrated in Figure\,\ref{fig:aliasingdiagram}. The IQ coefficients determined for the direct signal do not result in good separation for the aliased signal, though this can potentially be improved with better analog path equalization.

\subsection{System Stability}

By design, the COMAP system should be be very stable over timescales of minutes apart from fundamental known physical effects, particularly ``1/f'' amplifier gain fluctuations \citep{escotte_generation-recombination_2013, weinreb_multiplicative_2014} and atmospheric opacity variations. However several other effects were exposed during commissioning and routine observing.

Some of the IF amplifiers used proved to have unreliable bonding to the connectors and significant changes in the signal amplitude as a function of time were noted. These were more significant at the lower frequencies consistent with variable capacitive coupling of a broken connection. Replacing affected amplifiers resolved this issue.

Some of the LNAs exhibited changes in gain under certain circumstances. One of the symptoms was a change in the bias current when the calibration vane was inserted. This was ascribed to a change in power at the second stage that was rectified in the gate, changing the bias voltage. Choosing a different gate bias was sufficient to eliminate this. Another change in bias was associated with the pointing direction of the antenna and hypothesized to be due to light affecting the first stage of the LNA. Testing suggested that both visible and infrared light were implicated. A thin black polyethylene sheet was added behind the foam weather window, and an inserting a expanded polystyrene disc between the feed horns and the polyethylene to attenuate the infrared.

Ripple in the spectra were sometimes seen on windy days at certain angles of incidence of the wind on the receiver. The period corresponded to a standing wave between the weather window and the LNA input. The polystyrene foam disk mentioned above was modified to have a slightly convex outer surface that would put the weather window under tension and stabilize it. Note that it is the stability of the standing wave that matters, rather than merely its presence.

During early tests variations in the power measured in the spectrometer without corresponding changes in the powers monitored in the DCM2s. This appeared to be correlated with temperature changes in the digital rack, and lab tests confirmed that the ADC gains are temperature sensitive. The measurement indicated a gain sensitivity of $\sim-0.34$ \% \degr C, roughly twice the specification in the data sheet. Better temperature regulation was obtained after maintenance work on the air conditioning unit and relocation of its temperature sensor.

The Allan variance \citep{allan_statistics_1966} of the time series for each channel from the spectrometer quantifies how long an integration is consistent with statistical noise. For a series of consecutive channel power integrations, each of length $\tau$, the Allan variance is
\begin{equation}
    \sigma_{\mathrm{av}}^{2}=\frac{1}{2} \langle \left( \overline{p}_{\mathrm{n}+1} - \overline{p}_{\mathrm{n}} \right)^2 \rangle.
\end{equation}
In an ideal statistical noise limited system,
\begin{equation}
    \sigma_{\mathrm{av}}^{2} = \frac{1}{\mathrm{\Delta}f \tau},
\end{equation}
but generally, as $\tau$ tends to longer times, the Allan variance will assume larger values due to, for example, $1/f$ noise. If $\tau$ is less than this inflection point, integrations may be summed to give close to theoretical noise.

\begin{figure}
     \centering
         \includegraphics[width=0.5\textwidth]{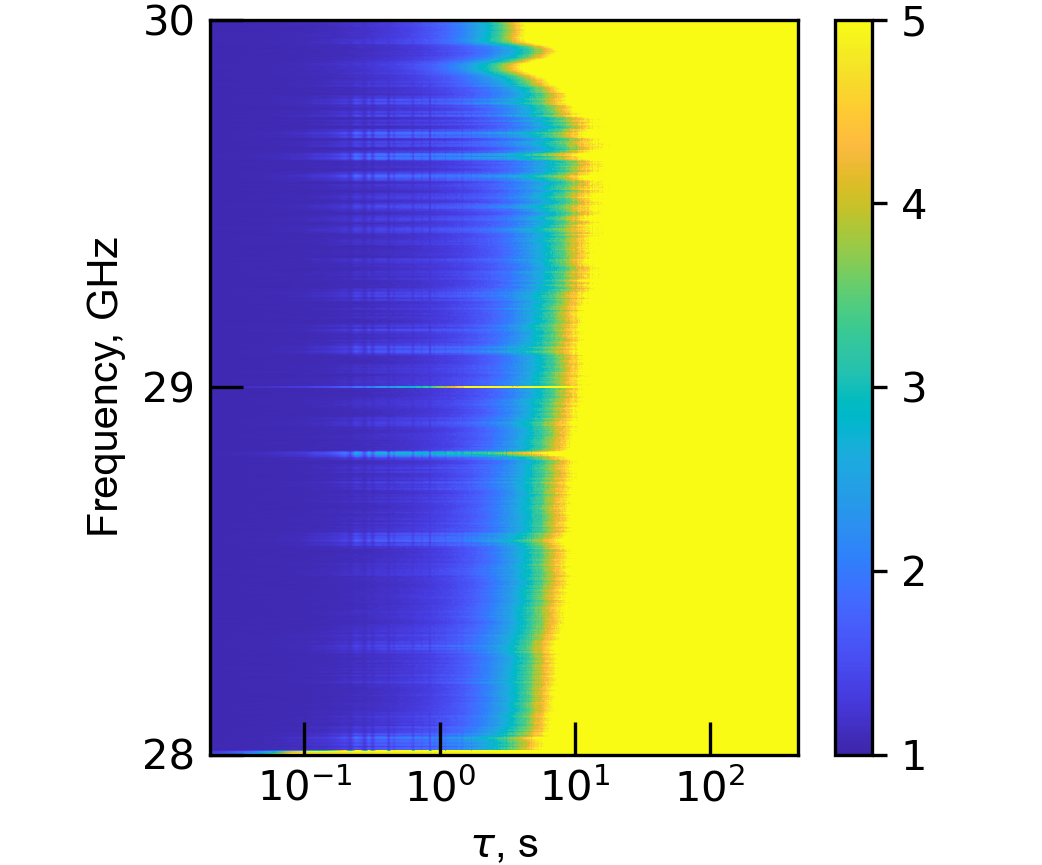}
         \includegraphics[width=0.5\textwidth]{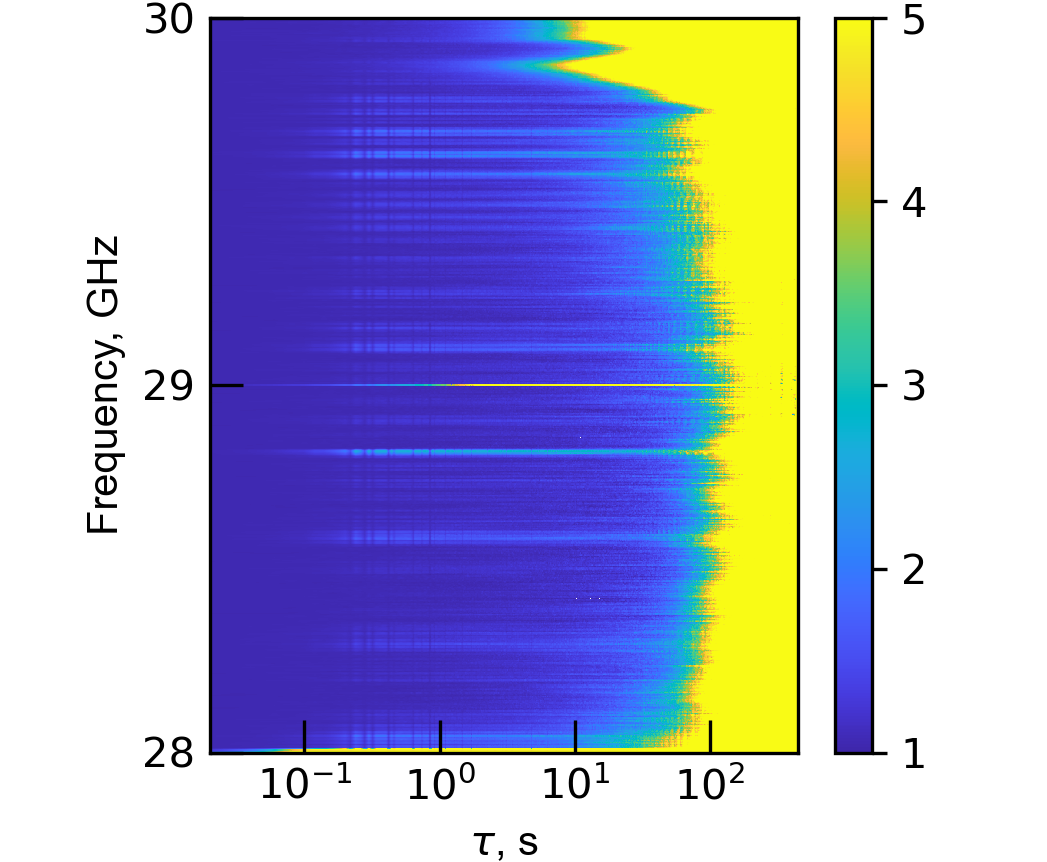}
        \caption{Allan variance normalized to $1/\Delta f \tau$ for the data from feed 1, band A, USB for a ten minute stare at zenith. \textit{Upper panel}:  values for the raw data stream. \textit{Lower panel}:  variance  calculated after removing a common-mode value from all channels.}
        \label{fig:avar}
\end{figure}

Figure\,\ref{fig:avar} shows the Allan variance on a range of time scales for all channels of one spectrometer. 
In Figure\,\ref{fig:avar} (upper panel) no processing has been applied to the data and it is apparent that it is not possible to integrate more than a few seconds before the sensitivity starts being compromised. The plot contains no information on how correlated the channels are, but subtracting the mean value of all channels at each time step, any common mode is removed. Figure\,\ref{fig:avar} (lower panel) shows that doing this pushes out the usable integration time up by about an order of magnitude. The common mode is an expected consequence of the gain fluctuations of the LNAs (quantified by a single parameter given as the product of the transistor mutual gains), and the atmospheric brightness changes. The gain fluctuations are common to all four bands while the atmospheric contributions have a significant commonality among feeds.

This very simple data processing indicates that scans over the source on time scales of $\sim 20$\,s can potentially result in statistical noise limited measurements. Results presented in the data pipeline and science results companion papers use much more sophisticated procedures that mitigate a variety of systematic effects.

%\newpage
\subsection{Interference} \label{subsec:rfi}

Generally radio frequency interference from coherent sources such as satellites and radio links, is not a significant problem at the observing frequency of COMAP. There are relatively few such sources, and the high directivity of the telescope minimizes pickup. Incoherent sources, such as power line arcing and ignition noise, are too weak to be detected at such high frequencies. Certain models of cell phone were found to produce noticeable effects when operated in the sidecab, but this is easily avoided.

During operation, certain broadband spikes were encountered in the data at certain times of year. These were found to be due to birds flying in front of the telescope aperture, or in front of the feeds. A different signature was seen from insects crossing the feed array, and the time resolution of the data enabled mapping their flight path over the pixels. Both the bird and insect signals vary significantly from year to year and with season, and were infrequent and obvious enough to flag with custom spike detection code.

The Sun and the Moon were also found to contribute fluctuations to the detected power. As described in Sec. \ref{subsec:ant}, there are some wide angle sidelobes resulting from scattering by the secondary support struts. If the Sun or Moon are present in those sidelobes the data are compromised, so data are flagged when either is closer to the main beam than about 60\degr.

\subsection{Verification on Astronomical Sources} \label{obs}

\subsubsection{Jupiter Observations} \label{jupiter}

\begin{figure}
    \centering
    \includegraphics{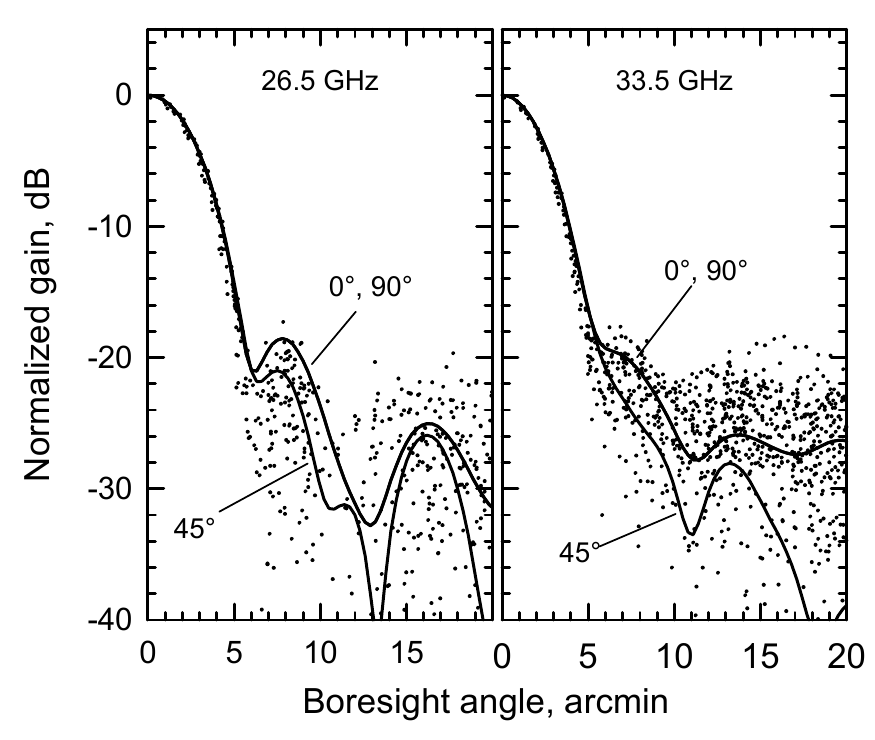}
    \caption{Measured and computed beam patterns for feed 1. The dots represent the measurements of the beam over all azimuthal angles round the boresight direction. Data are averaged over 1\,GHz bandwidth. Solid lines show the computed patterns at azimuth angles of 0\degr \ , 45\degr \ and 90\degr.}
    \label{fig:beampat}
\end{figure}

Regular observations are made of Jupiter to verify performance and check pointing. These are constant elevation scans that require different processing than the CO fields which suppress continuum sources. Representative beam patterns (Figure\,\ref{fig:beampat}) agree well with predictions, showing the modest change in beamwidth with frequency.

Pointing and beam shapes are confirmed for all feeds. Figure\,\ref{fig:beammaps2d} plots the beams in the nominal coordinates predicted for their pointing centers. Deviations from those are it the limit of the measurement accuracy.

\subsubsection{Beam Maps} \label{subsubsec:beammaps}
\begin{figure}
    \centering
    \includegraphics[width=0.425\textwidth]{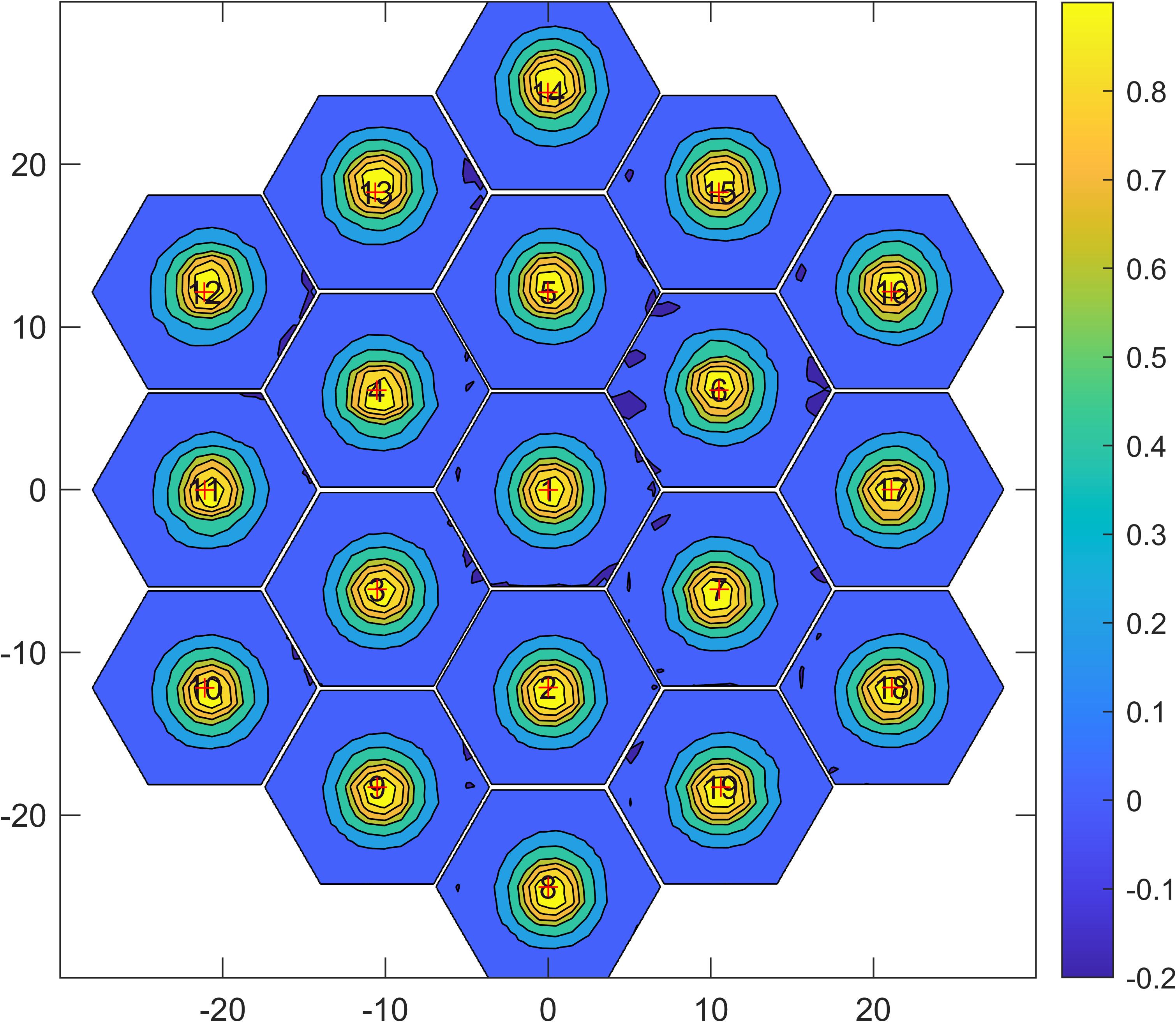}
    \caption{Beam maps derived for each feed from raster scans on Jupiter. Note that these are reversed left to right and top to bottom compared to a map of Jupiter since, e.g., pointing above Jupiter is a measure of a point in the beam below boresight.}
    \label{fig:beammaps2d}
\end{figure}

Verification of the beam shape and offset using Jupiter as an unresolved source are in good accord with the predictions. The maps for all the feeds are shown in Figure\,\ref{fig:beammaps2d}. 

\subsubsection{Spectral Line Observations} \label{subsec:specline}
\begin{figure}
    \centering
    \includegraphics{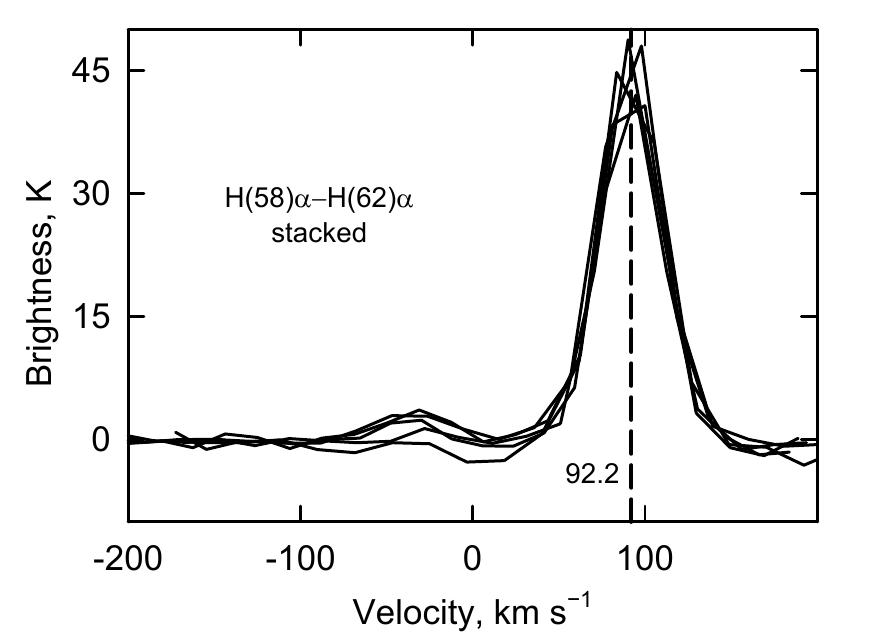}.
    \caption{Radio recombination lines observed in the COMAP passband in the star-forming region W43. The vertical dashed line indicates the expected line center due to the known velocity of the region.}
    \label{fig:recomb}
\end{figure}

A project to map significant portions of the Galactic plane has yielded significant detections of several H$\alpha$ radio recombination lines (Figure\,\ref{fig:recomb}), and potential detections of C and He recombination lines \citep{es_VI}. These serve as an important confirmation of the mapping between ROACH-2 channels and sky frequency. Additionally, they demonstrate that the instrumental spectral baselines are well enough controlled to produce high-significance line measurements.

\section{Discussion} \label{sec:discussion}

The COMAP Pathfinder was conceived as a proof-of-concept for a line intensity instrument that would allow exploration of the challenges posed by deep integration with a total power system. It was also expected to place constraints on the range of models of CO distribution at redshifts $z = 2.4$—$3.4$. The rapid deployment was made possible by the availability of a high-performance 10.4\,m antenna. With a modification of the optics, including a new secondary mirror, it was well suited to a 19 feed 26--34\,GHz receiver.

Design efforts were focused on the stability of the gain, particularly as a function of frequency, on timescales of a couple of tens of seconds. Amplifier gain instabilities are tied to the fundamental physics of the transistors, but since they appear as a common mode across the band they can be filtered out. Similarly, fluctuations of the atmospheric brightness are very smooth in frequency and have a strong common mode component among the feeds. Most of the problematic changes are related to standing waves in transmission paths. The data analysis is insensitive to ripple due to standing waves, but very sensitive to changes in the ripple, typically caused by temperature changes or mechanical stress from the antenna motion.

Standing waves in the optics are particularly troublesome since they are at the highest frequencies. By using circular polarization standing waves are greatly reduced. During the first observing season two feeds had no polarizers, two had single-section polarizers and the remainder had two-section polarizers. An unanticipated consequence was that interactions between the polarizers and corrugated feeds caused strong noise spikes in the band. However, the overall improvement in the data quality from the analysis pipeline with the polarizers installed outweighed the loss in discarding the affected frequency channels. All feeds now have the two stage polarizers. Further development will include eliminating the spikes while retaining the polarizers.

Ripple due to a standing wave between the LNAs and the weather window was found to cause problems under certain conditions of wind and antenna orientation. This was easily solved by adding a foam backing to hold the window taught.

With careful selection of cable types, good temperature regulation, and avoidance of mechanical stress, Cable standing waves proved not to be a limitation. In certain instances where cable standing waves were identified, it was traced back to a break in a connector or component that was the fixed or replaced.

The most concerning limitation of the system was revealed to be pickup from ground spillover and bright astronomical sources. While this was anticipated, it was not easy to quantify in advance. With appropriate scanning patterns and avoiding bright sources such as the Sun, good data are still achievable. However, for future experiments much of this type of contamination can be avoided with a dual-offset antenna. Mechanically, offset antennas are more massive to achieve the same stiffness, and have a larger moment of inertia and therefore higher drive requirements. For the Pathfinder, the choice of an existing, well-understood, high performance antenna was a critical factor in getting good quality data in on a short timescale that significantly outweighed potential benefits of a new unevaluated offset antenna.

Key requirements for the instrument have been informed by continuing theoretical simulations \citep{es_V,es_VII}, while the pipeline analysis \citep{es_III}, and science data processing \citep{es_IV,es_VI} have been crucial in evaluating performance and detecting problems and verifying fixes. Since the first season of observations several improvements have been made, such as upgrading LNA modules, installing two-stage polarizers on all feeds, and repairing faulty components. So far, no fundamental limits to this method of line intensity mapping have been encountered in the first season of observing, and the current and imminent modifications will improve the data quality for future season. Future instruments, such as a complementary 16\,GHz system \citep{es_I}, will benefit greatly from the experience with the COMAP Pathfinder.

\newpage
\vspace{5mm}
\section*{Acknowledgments}
This material is based upon work supported by the National Science Foundation under Grant Nos.\ 1517108, 1517288, 1517598, 1518282 and 1910999, and by the Keck Institute for Space Studies under ``The First Billion Years: A Technical Development Program for Spectral Line Observations''. Parts of the work were carried out at the Jet Propulsion Laboratory, California Institute of Technology, under a contract with the National Aeronautics and Space Administration, and funded through the internal Research and Technology Development program. DTC is supported by a CITA/Dunlap Institute postdoctoral fellowship. The Dunlap Institute is funded through an endowment established by the David Dunlap family and the University of Toronto. CD and SH acknowledge support from an STFC Consolidated Grant (ST/P000649/1). JB, HKE, MKF, HTI, JGSL, MR, NOS, DW, and IKW acknowledge support from the Research Council of Norway through grants 251328 and 274990, and from the European Research Council (ERC) under the Horizon 2020 Research and Innovation Program (Grant agreement No.\ 819478, \textsc{Cosmoglobe}). 
JG acknowledges support from the University of Miami and is grateful to Hugh Medrano for assistance with cryostat design. 
LK was supported by the European Union’s Horizon 2020 research and innovation program under the Marie Skłodowska-Curie grant agreement No.\ 885990. 
J.\ Kim is supported by a Robert A.\ Millikan Fellowship from Caltech. 
At JPL, we are grateful to Mary Soria for assembly work on the amplifier modules and to Jose Velasco, Ezra Long and Jim Bowen for the use of their amplifier test facilities. 
HP acknowledges support from the Swiss National Science Foundation through Ambizione Grant PZ00P2\_179934. PCB is supported by the James Arthur Postdoctoral Fellowship. RR acknowledges support from ANID-FONDECYT grant 1181620. MV acknowledges support from the Kavli Institute for Particle Astrophysics and Cosmology. We thank Isu Ravi for her contributions to the warm electronics and antenna drive characterization. We would also like to acknowledge the important contributions of Russell Keeney, Michael Virgin, and Andres Rizo for constructing and installing major mechanical parts of the instrument

\appendix \label{sec:appendix}
\section{Design Principles}
As a total power instrument, COMAP must discriminate a tiny signal from a system noise level that is about six orders of magnitude greater. This system noise results from the instrument (antenna and receiver), the environment (ground and atmosphere, radio interference), and foreground astronomical sources (diffuse and compact). The key goal was to build a system that is stable over times comparable to or longer than the time to scan a source. Faster scanning allows the source to be covered in a shorter time, but particularly for an existing telescope there is a limit to the maximum rate. For the 10.4\,m Leighton antenna, the typical time to scan over a degree-sized patch is a few seconds. Ideally, fluctuations in the total power should be limited by the statistics of the radiometer (statistical) noise. In practice, there are some background fluctuations that are common to all frequencies or pixels that can be easily removed, but other more subtle artifacts can easily dominate the data.

Most phenomena fall into one of two categories. These are essentially multiplicative (gain and passband changes), or additive (ground pickup, bright astronomical sources. For COMAP, these are dominated by standing waves and sidelobes, respectively.

\subsection{Standing Waves} \label{subsec:sw}

Frequency-dependent variations in the output spectra of the instrument are many orders of magnitude greater than the deviations due to the expected CO signal and cannot be determined a priori to an acceptable accuracy ($\lesssim 10^{-7}$). The structure is primarily a result of the passband shape of the receiver and optics, and to a lesser extent to variations of noise temperature with frequency. Details of how the data analysis deals with distinguishing the spatial and temporal structure of the source from the instrument and foregrounds are discussed by \citet{es_III}. For the purposes of this paper, we can summarize the goal as reducing the temporal systematic variations in the spectrometer output to a level comparable to or less than the statistical noise over a time of order 10--30 seconds after simple temporal and frequency slopes have been removed. Provided the systematic residuals are random among different observations and feeds, they will not add coherently and will remain low compared to the statistical noise.

Ripple in transmission paths is a dominant source of spectral structure, so stabilizing transmission path lengths was a major focus in the instrument design. Standing waves are produced by discontinuities between sections of a transmission path. Multiple standing waves generate periodic ripple in the passband of a system, with a periodicity related to the time delay of wave transmission between pairs of discontinuities. Any pair of discontinuities give rise to ripple that will correspond to a particular value in the $k$-spectrum\footnote{Note that a standing wave produced by a \textit{single} discontinuity does not result in ripple in the passband.}. For a lossless transmission line with a length $l$, and small voltage reflection coefficients $r_1$ and $r_2$ at the two ends, the complex amplitude transmission function is approximately
\begin{equation}
    h(\omega) = (1-r_1 r_2 )(1+r_1 r_2 e^{\mathrm{i} \omega l / 2 n c}),
\end{equation}
where $\omega$ is the angular frequency of the wave, $c$ is the speed of light in vacuum, and $n$ is the propagation velocity in the medium relative to vacuum. When applied to a white noise spectrum with unit spectral density, the output spectrum has a sinusoidal ripple given by
\begin{equation}
\begin{aligned}
P(\omega)& = P_\mathrm{r} \cos {\omega \tau}, \\
P_\mathrm{r} & = 2(1-r_1 r_2)^2 r_1 r_2,\\
P_0 &= 2(1-r_1 r_2)^2 (r_1^2 r_2^2+1),\\
\tau &= l/2nc,
\end{aligned}
\end{equation}
where we assume without much loss of generality that the reflection coefficients are real. In the data processing every frequency channel is initially normalized against its running mean, so if this ripple in the power is constant in time it does not contaminate the astrophysical power spectrum. Commonly, the reflection coefficients are relatively stable if proper connections are made, but the effective length of the line can vary on different time scales due to temperature or flexing, for example, causing instability in the ripple pattern. Treating the reflection coefficients as small, and assuming an incremental change in the time delay of the line of $\mathrm{\delta}\tau$, we arrive at a residual ripple in the power spectrum given by
\begin{align}
\delta P(\omega)& = P_\mathrm{r} \mathrm{\delta} \tau \omega \sin {\omega \tau}, &
\end{align}
Assume that we take a series of spectra over a time $t_\mathrm{scan}$, and that $\tau$ changes linearly over that time by an amount $\delta \tau$. For each of the integrated spectra  we can calculate the standard deviation due to the ripple across each spectrum and then compute the mean standard deviation over all spectra. For small changes in the line length, and some reasonable approximations, the average standard deviation can be written as
\begin{equation}
    \sigma_r = \frac{\pi}{\sqrt{2}} \mathrm{\delta} \tau P_\mathrm{r} f_\mathrm{mean},
\end{equation}
where $f_\mathrm{mean}$ is the mean frequency across the spectrometer band.

It is evident that the effect is more significant for higher frequencies. Standing waves in the optics at 26--34\,GHz are an order of magnitude more sensitive to length changes than are the standing waves in the cables to the spectrometers at 0--2\,GHz.

Comparing the standard deviation of the ripple with the statistical noise aids understanding of the stability requirements. If each spectrum has a channel bandwidth $\Delta f$ and is integrated for a time $t_\mathrm{int}$, then the standard deviation of the power across the spectrum is
\begin{equation}
    \sigma_\mathrm{stat} = \frac{1}{\sqrt{\Delta f t_\mathrm{int}}}.
\end{equation}

If we require $\sigma_\mathrm{r} \ll \sigma_\mathrm{stat}$ over a scan time $t_\mathrm{scan}$ and the spectra are integrated over a time $t_{\mathrm{int}}$, then the rate of change of the line length should ideally be limited to
\begin{equation}\label{eq:rateofchange}
\frac{\mathrm {d} \tau} {\mathrm {d} t} \ll \frac {2}{\pi f_\mathrm{mean} P_{\mathrm {r}} t_\mathrm{scan} t_{\mathrm{int}}^{1/2} \Delta f^{1/2}}.
\end{equation}
Typically, this criterion may be re-written as a rate of temperature change, since temperature is the main driver of the line length, $l$. Quantitative examples are given in the main text. 

Particular regard was paid to the optical standing wave because of the sensitivity at the high frequencies as noted above. The three main strategies applied to mitigate this were: use of circular polarization; implementation of a diverter cone at the secondary; and thermal compensation of the receiver support. Significant suppression of the optical standing wave ripple is achieved by using circular polarization as detailed in \ref{subsubsec:wgcmpts}.

In the rest of the system, LMR cable (Times Microwave Systems) was used in most of the signal paths for several reasons. The dielectric is polyethylene, which does not suffer from the ‘knee’ in the delay vs. temperature curve around 20°C observed in the more common PTFE \citep{czuba_temperature_2011}, which would make the standing waves very susceptible to temperature changes. LMR cable (apart from the smallest diameter) uses foamed polyethylene which reduces the loss, as well as reducing temperature sensitivity further. Any standing wave corresponds to a particular $k$-value. It was not possible to make all the cables short or long enough to place their ripple period out of the COMAP $k$-value range, though that was done where possible.

\subsection{Antenna Spillover and Sidelobes} \label{subsec:antspill}

Another focus in the design was control of the spillover of the telescope. Radiation entering from the ground and strong celestial sources is a particular problem because it varies with the telescope pointing making it difficult to separate from the target signal. Since the telescope chosen for COMAP is a symmetric Cassegrain, there are limits on scattering due to unavoidable blockage by the secondary and its support struts.

Any parts of the antenna radiation pattern\footnote{For simplicity, we treat the antenna as a transmitter; reciprocity dictates that the receiving gain pattern is identical to the transmitting pattern.} not covering the target field have the potential to add contaminating radiation from other sources. Primarily, these are the Sun, the Moon and the Earth. Suppression below the signal level requires the pattern integrated across the source to be more than 80 dB below the integrated main beam. For the COMAP there are sidelobes high enough that the Sun and Moon significantly exceed this, and this has to be taken account of in the observing strategy. Ground radiation, while having a lower brightness temperature than the Sun, covers a much greater solid angle and therefore is difficult to avoid. Primarily it enters through fields diffracted at the secondary mirror perimeter that spill past the primary. The optical design was targeted at minimizing ground spillover, as discussed in section 3.3.1.

\bibliography{COMAP_instrument_library,early_science,online.bib}{}
\bibliographystyle{aasjournal}

\end{document}